\documentclass[reprint, aps,prd,twocolumn,showpacs,showkeys, 10pt, longbibliography, nolinenumbers, floatfix]{revtex4-2}

\usepackage{amsmath}
\usepackage{graphicx}
\usepackage{float}
\usepackage{dcolumn}
\usepackage{multirow}
\usepackage{natbib}
\usepackage{changes}
\usepackage{amsmath}
\usepackage[colorlinks = true,linkcolor = blue,urlcolor  = blue,citecolor = blue,anchorcolor = blue]{hyperref}
\usepackage{enumerate}
\usepackage{tabularray}   
\usepackage{lipsum}  
\usepackage{comment}
\usepackage{times}
\usepackage{microtype}

\begin{document}

\title{Hybrid stars in light of the HESS J1731-347 remnant and the PREX-II experiment}

\author{P. Laskos-Patkos}
\email{plaskos@physics.auth.gr}
\author{P.S. Koliogiannis}
\email{pkoliogi@physics.auth.gr}
\author{Ch.C. Moustakidis}
\email{moustaki@auth.gr}

\affiliation{Department of Theoretical Physics, Aristotle University of Thessaloniki, 54124 Thessaloniki, Greece}

\begin{abstract}
The recent analysis on the central compact object in the HESS J1731-347 remnant suggests interestingly small values for its mass and radius. Such an observation favors soft nuclear models that may be challenged by the observation of massive compact stars. In contrast, the recent PREX-II experiment, concerning the neutron skin thickness of $^{208}$Pb, points toward stiff equations of state that favor larger compact star radii. In the present study, we aim to explore the compatibility between stiff hadronic equations of state (favored by PREX-II) and the HESS J1731-347 remnant in the context of hybrid stars. For the construction of hybrid equations of state we use three widely employed Skyrme models combined with the well-known vector MIT bag model. Furthermore we consider two different scenarios concerning the energy density of the bag. In the first case, that of a constant bag parameter, we find that the resulting hybrid equations of state (which satisfy the HESS J1731-347 constraints) are strongly disfavored by the observation of $\sim2 M_\odot$ pulsars. However, the introduction of a Gaussian density dependence yields results that are compatible with the conservative $2 M_\odot$ constraint. The utilization of recent data based on the observation of PSR J0030+0451, PSR J0952-0607 and GW190814 allows for the imposition of additional constraints on the relevant parameters and the stiffness of the two phases. Interestingly, we find that the derived hybrid equations of state do not satisfy the PSR J0030+0451 constraints in $1\sigma$ and only marginally agree with the $2\sigma$ estimations. In addition, we estimate that the observation of massive pulsars, like PSR~J0952-0607, in combination with the existence of HESS J1731-347, may require a strong phase transition below $\sim 1.7n_0$. Finally, we show that the supermassive compact object ($2.5$-$2.67M_\odot$) involved in GW190814 could potentially be explained as a rapidly rotating hybrid star.

\end{abstract}
\maketitle
%%%%%%%%%%%%%%%%%%%%%%%%%%
\section{Introduction} \label{section1}
%%%%%%%%%%%%%%%%%%%%%%%%%%%
One of the most important unresolved questions regarding the physics of dense nuclear matter concerns the density dependence of the nuclear symmetry energy~$E_{sym}(n)$~\cite{Baldo-2016,Roca-Maza-2018}. A critical parameter for the behavior of isospin asymmetric matter is the so-called symmetry energy slope $L$, which is proportional to the derivative of $E_{sym}$ at the saturation density.
Despite its significance, this quantity cannot be directly measured in experiments. Therefore, the identification and  use of relevant observables on finite nuclei are essential for its accurate determination. In that direction, Reed {\it et al.}~\cite{Reed-2021} employed the recent PREX-II estimation on the neutron skin thickness of $^{208}$Pb ($\Delta r_{np}=0.283 \pm 0.071 $ fm)~\cite{Abrahamyan-2012,Horowitz-2012,Adhikari-2021} and exploited its strong and linear correlation with the slope parameter to report a value of $L=106 \pm 37$ MeV. Interestingly, the latter result is  higher than previous theoretical calculations or theoretical interpretations of experimental data~\cite{Hebeler-2013,Zhang-2013,Hagen-2016,Drischler-2020,Horowitz-2014}, and points toward a rather stiff equation of state (EOS). From an astrophysical perspective, the aforementioned $L$ value would result in compact star configurations with large radii and tidal deformabilities~\cite{Reed-2021}. Recently, the CREX Collaboration published their data on the neutron skin thickness of $^{48}$Ca and they reported a very low value of $\Delta r_{np}=0.121 \pm 0.026 $~fm~\cite{Adhikari-2022}. The latter result favors soft EOSs and hence it is in tension with the PREX-II results~\cite{Reinhard-2022,Tagami-2022}. It is worth mentioning that several studies until this moment, have attempted and failed to resolve this discrepancy~\cite{mKumar-2023,Miyatsu-2023}. Notably, the authors of Ref.~\cite{Reed-2023} achieved the construction of three energy density functionals (EDFs) that can simultaneously describe the PREX and CREX measurements. However, the resulting large and positive values for the curvature of the symmetry energy ($K_{sym}$) are in striking contrast to most nonrelativistic EDFs and {\it ab initio} models that predict negative values~\cite{Reed-2023}. Consequently, the derived EOSs are extremely stiff. The upcoming Meinz radius
experiment at MESA~\cite{Becker-2018} will either verify or disprove the PREX-II results, and it will hopefully shed light on the PREX-CREX tension.

The PREX-II estimation for the stiffness of the EOS may have rather interesting astrophysical implications. For instance, in Ref.~\cite{Reed-2021}, the authors used a special class of EDFs and argued that the radius of a $\sim1.4M_\odot$ neutron star ($R_{1.4}$) should be at least $\sim13$km, while for the dimensionless tidal deformability they predicted a minimum value of $\Lambda_{1.4}\sim642$. Notably, according to the analysis of Ref.~\cite{Abbott-2018}, the aforementioned $\Lambda_{1.4}$ estimation is not compatible with the GW170817 event~\cite{Abbott-2017}. In a recent study, Thakur {\it et al.}~\cite{Thakur-2023} employed a relativistic mean field approach and predicted a similar value for $R_{1.4}$, where the resulting tidal deformability marginally satisfied the constraint from Ref.~\cite{Abbott-2018}. It is worth commenting that, several different studies, which employed recent astronomical data, have also found difficulty in accommodating the PREX-II values of $\Delta r_{np}$ and $L$~\cite{Essick-2021a,Essick-2021b,Yeunhwan-2022,Thapa-2023,Soares-2023}. In that direction, it has been suggested that a possible phase transition from hadronic to deconfined quark matter may be essential to ensure the stiff EOS behavior at low baryon densities and a subsequent softening at higher density values (favored by GW170817)~\cite{Chen-2023}. Finally, the PREX-II result has also an intriguing impact on the thermal evolution of compact stars. In particular, a large symmetry energy slope lowers the stellar mass threshold for the activation of the direct Urca process, leading to rapid cooling~\cite{Thapa-2022,Sarkar-2023}.

A recent analysis on the central compact object within the HESS J1731-347 supernova remnant indicates that it has an interestingly small mass and radius~\cite{Doroshenko-2022}. Specifically, within 1$\sigma$, $M=0.77^{+0.20}_{-0.17}M_\odot$ and $R=10.4^{+0.86}_{-0.78}$km. It is worth mentioning that, as Alford and Halpern~\cite{Alford-2023} commented, the aforementioned result relies on the assumption that the star has a uniform temperature carbon atmosphere and it is located at a distance of 2.5 kpc. As a consequence, further studies are required to cross examine and understand the validity of these results In any case, the possibility of such a strangely light compact star has triggered a wave of studies that investigate its nature.

To begin with, Di Clemente {\it et al.}~\cite{DiClemente-2023} attempted to explain the formation of the aforementioned compact object in the framework of strange stars. In particular, the authors argued that strange stars exhibit lower gravitational mass for a given baryon mass. Hence, despite the fact that a recent analysis points to a minimum neutron star formation mass of $1.17 M_\odot$ (based on the estimation of a minimum baryon mass of 1.28 $M_\odot$)~\cite{Suwa-2018}, this value could be lower in the presence of strange quark matter. Later, Horvath {\it et al.}~\cite{Horvath-2023} examined the case of a strange star by adopting a model of color-flavor locked quark matter. Apart from the mass-radius dependence, the authors also discussed the thermal evolution and argued that the suppression of rapid cooling due to superconductivity may suffice for the understanding of the reported temperature and age of the HESS~J1731-347 remnant. The strange star scenario has also been studied in Refs.~\cite{Oikonomou-2023,Das-2023,Rather-2023} with the implementation of additional constraints (trace anomaly, GW170817) and the discussion of different stellar properties (e.g., anisotropy, radial oscillations). 

An investigation has also been conducted  for the scenario of hybrid stars~\cite{Tsaloukidis-2023,Brodie-2023,Sagun-2023}.~The systematic study of Ref.~\cite{Tsaloukidis-2023} on the properties of twin stars showed that, depending on the transition density and the hadronic EOS, a third family of compact objects could also provide an explanation for the mass and radius reported in Ref.~\cite{Doroshenko-2022}. Furthermore, Brodie and Haber~\cite{Brodie-2023} constructed a large set of EOSs based on the results of chiral effective field theory and  showed that purely hadronic stars barely cross the $2\sigma$ contour for the compact object in HESS J1731-347. However, by combining the softest hadronic EOS from their set with a constant speed of sound quark model, the authors concluded that hybrid stars may provide a viable explanation. In addition, the recent study of Ref.~\cite{Sagun-2023}, examined the cooling aspect of the problem and showed that hybrid stars are potentially compatible with the thermal evolution of the HESS~J1731-347 remnant.

It is important to comment that, according to Refs.~\cite{Sagun-2023,Huang-2023,Li-2023,Kubis-2023}, the case of a purely hadronic star should not be excluded. However, the common conclusion from the aforementioned studies is that the EOS needs to be quite soft in order to reproduce the results of Ref.~\cite{Doroshenko-2022} in $1\sigma$. Notably, the authors of Ref.~\cite{Routaray-2023} have managed to provide a description for the HESS~J1731-347 remnant by employing a rather stiff EOS and by considering the admixture of dark matter. While the original EOS fails to explain the reported radius, a proper value for the Fermi momentum of dark matter yields the desired result.

The main motivation of the present study is twofold. Firstly, we aim to explore the compatibility between stiff hadronic EOSs (favored by PREX-II) with the observation of the HESS J1731-347 remnant in the context of hybrid stars. Secondly, we wish to examine the properties of the resulting hybrid stars in the light of recent astrophysical constraints.  Interestingly, almost none of the EOSs that were previously employed for the explanation of the compact object in HESS J1731-347 (within 1$\sigma$) is compatible with the results of PREX-II. In particular, the only EOS that lies within the PREX-II range is employed in Ref.~\cite{Sagun-2023}. It is worth noting that, despite the relatively large symmetry energy slope, the authors refer to that EOS as soft due to its low incompressibility value of $K=201$ MeV. 

This paper is organized as it follows. In Sec.~\ref{section2} we review the hadronic and quark models selected in this work. In addition, we briefly discuss Maxwell construction, which is the employed method for the derivation of hybrid EOSs. In Sec.~\ref{section3} we present our results and discuss their implications. Section~\ref{section4} contains a summary of our findings.

\section{Equation of State} \label{section2}
\subsection{Hadronic matter}
For the description of the hadronic part we employed three distinct Skyrme effective interactions~\cite{Kohler-1976,Reinhard-1995}. In particular, we use the Ska, SkI3 and SkI5 EOSs. The corresponding cluster energy functional used for the unified description of the inner crust is the one from Ref.~\cite{Danielewicz-2009}. In addition, for the outer parts of stellar models we have added the well-known EOS of Baym {\it et al.}~\cite{Baym-1971}. It is worth noting that all of the employed hadronic EOSs have been extensively used in the literature for the study of cold and hot neutron star matter~\cite{Chabanat-1997,Costantinou-2014,Costantinou-2015,Tsaloukidis-2019,Margaritis-2021,Biswas-2022,Landry-2022,Mikheev-2023}.

\begin{figure}
  \centering  \includegraphics[width=9 cm,scale=1]{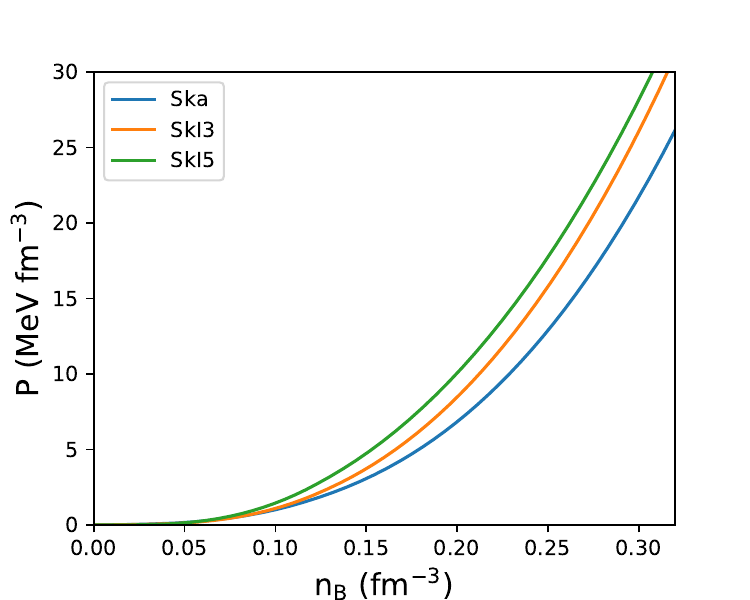}
  \caption{The hadronic EOSs employed in this study.}
  \label{Fig1}
\end{figure}

The key nuclear properties for the employed EOSs can be found in Table~\ref{tab:table1}. Notably, the corresponding values for the symmetry energy slope cover the wide interval predicted by the PREX-II Collaboration ($L=106\pm37$ MeV)~\cite{Reed-2021}. Furthermore, the generally accepted value for the incompressibility of nuclear matter is considered to be $K=240\pm20$~MeV~\cite{Shlomo-2006,Colo-2008,Garg-2018}. However, according to the comprehensive analysis of Ref.~\cite{Stone-2014}, the value of $K$ could be larger. For the latter reason the selected EOSs are all characterized by $K\sim260$ MeV, which corresponds to the upper limit of the aforementioned range. In Figure~\ref{Fig1}, we depict the considered hadronic EOSs. As it is evident, the larger the symmetry energy slope, the higher the pressure for a given baryon density (i.e., the stiffer the EOS).

The employed EOSs have been obtained through the compOSE repository~\cite{compose}. For more details concerning their construction the reader is referred to Refs.~\cite{Kohler-1976,Reinhard-1995,Danielewicz-2009,Gulminelli-2015}

\begin{table}[H]
\caption{\label{tab:table1}
The energy per nucleon of symmetric nuclear matter $E_0$, the symmetry energy slope $L$, the incompressibility $K$ and the symmetry energy $J_0$ at saturation density $n_0$ for the employed hadronic models. }
\begin{ruledtabular}
\begin{tabular}{cccccc}
Model & $n_0$ (fm$^{-3}$)& $E_0$ (MeV) &$L$ (MeV) & $K$ (MeV)& $J_0$ (MeV) \\
\hline
Ska & 0.155 & -15.99 & 74.62 & 263.2 & 32.91 \\
SkI3 & 0.158 & -15.98 & 100.5 & 258.2 & 34.83 \\
SkI5 & 0.156 & -15.85 & 129.3 & 255.8 & 36.64 \\
\end{tabular}
\end{ruledtabular}

\end{table}

\subsection{Quark matter}

The worldwide model for the description of quark matter is the MIT bag model~\cite{Baym-1976}. In this framework, noninteracting quarks are confined within a bag due to the enforcement of an external pressure, known as vacuum pressure. The corresponding Lagrangian density for the standard MIT bag model is the following~($\hbar=c=1$)~\cite{Lopes-2021a}
\begin{equation} \label{eq1}
    \mathcal{L}_0=\sum_{q=u,d,s} [\Bar{\psi_q} (i\gamma_{\mu}\partial^{\mu}-m_q)\psi_q-B]\Theta,
\end{equation}
where $m_q$ stands for the rest mass of q quark and $B$ is the so-called bag constant. $\Theta$ denotes a Heavyside step function which vanishes out of the bag in order to ensure the confinement.

As performed in multiple previous studies~\cite{Lopes-2021a,Lopes-2021b,Klahn-2015,Gomez-2016,Gomez-2019a,Gomes-2019b,Jaikumar-2021,Costantinou-2021,Zhao-2022,Costantinou-2023,Lyra-2023,Kumar-2022,Kumar-2023}, we introduce an interaction among quarks via minimal coupling with a vector boson $V^\mu$, which is analogous to the $\omega$ meson in quantum hadrodynamics~\cite{Serot-1992}. Therefore, the interaction term reads as~\cite{Lopes-2021a}
\begin{equation} \label{eq2}
    \mathcal{L}_{\rm vec}=\{-g_v\sum_{q} \Bar{\psi_q} \gamma_\mu V^\mu\psi_q + \frac{1}{2}m_{V}^2 V_\mu V^\mu\}\Theta,
\end{equation}
where $g_v$ is the coupling constant and $m_{V}$ denotes the mass of the mediating boson.~In the present model, known as the vector MIT Bag model, the energy density can be explicitly written  as~\cite{Jaikumar-2021}
\begin{equation} \label{eq3}
    \mathcal{E}_{Q} = \sum_{q} \mathcal{E}_q + \frac{1}{2} G_v (n_u+n_d+n_s)^2+B,
\end{equation}
where $G_{v}=(g_{v}/m_{ V})^2$. For simplicity, we will refer to $G_v$ as a coupling constant since it incorporates the strength of the repulsive interaction. The quantity $\mathcal{E}_q$ corresponds to the energy density of an ideal Fermi gas, given as
\begin{equation}  \label{eq4}
\begin{split}
    \mathcal{E}_q  & =  \frac{3}{8\pi^2}  \Biggl\{ k_{Fq}(k_{Fq}^2+m_q^2)^{1/2}(2k_{Fq}^2+m_q^2)   \\ 
    & -   m_q^4\ln \left [ \frac{k_{Fq}+(k_{Fq}^2+m_q)^{1/2}}{m_q}\right]  \Biggl\}, 
\end{split}
\end{equation}
where $k_{Fq}=(\pi^2n_q)^{1/3}$ denotes the quark Fermi momentum. It is worth commenting that the formula of Eq.~(\ref{eq3}) could be equivalently derived by the consideration of a Yukawa type potential~\cite{Yang-2021,Yang-2023}. Furthermore, the inclusion of a vector interaction shifts the chemical potential of q quark as~\cite{Jaikumar-2021}
\begin{equation} \label{eq5}
    \mu_q=(k_{Fq}^2+m_q^2)^{1/2}+G_v(n_u+n_d+n_s),
\end{equation}
and the total pressure of the system is found through the standard thermodynamic relation
\begin{equation} \label{eq6}
    P_Q = \sum_{q} \mu_q n_q - \mathcal{E}_Q.
\end{equation}

At this point, there are 3 degrees of freedom ($n_u,n_d,n_s$) that determine the thermodynamic state of the system. In order to calculate the quark EOS we also need to consider the conditions for chemical equilibrium and charge neutrality. Taking into account the relevant weak processes, considering a system composed of uds quarks and electrons, the condition for chemical equilibrium reads as
\begin{equation} \label{eq7}
    \mu_d=\mu_u+\mu_e, \quad \mu_d=\mu_s,
\end{equation}
while the charge neutrality is expressed via
\begin{equation} \label{eq8}
    \frac{2}{3}n_u-\frac{1}{3}(n_d+n_s)-n_e=0.
\end{equation}
The quantities $\mu_e=((3\pi^2 n_e)^{2/3}+m_e^2)^{1/2}$, $n_e$ and $m_e$ denote the electron chemical potential, number density and mass, respectively.

\subsection{Phase transition}

In the present work we considered the scenario of a sharp phase transition. Namely, we employed Maxwell construction, which is the favored method in the case of a large surface tension in the hadron-quark interface ($\sigma\gtrsim40$ MeV fm$^{-2}$)~\cite{Yasutake-2016,Mariani-2017}. In this framework, both phases need to satisfy local charge neutrality, which results into a discontinuity in the energy density ($\Delta\mathcal{E}$). In principle, the favorable phase is characterized by the lower baryon chemical potential for a given value of pressure. Thus, the onset of quark deconfinement is determined by the following conditions~\cite{Bielich-2020}
\begin{equation}
    P^h=P^q, \quad \mu^h_B=\mu^q_B, \quad T^h=T^q,
\end{equation}
where $P$, $\mu_B$ and $T$ denote the pressure, the baryon chemical potential and the temperature, respectively. The superscripts $h$ and $q$ stand for the hadronic and the quark phase. Note that the condition for thermal equilibrium is trivially satisfied due to the use of zero temperature EOSs. Finally, the baryon chemical potential for hadronic matter is given by
    $\mu_B^h=\mu_n$,
where $\mu_n$ is the chemical potential of the neutron. For quark matter: $\mu_B^q=2\mu_d+\mu_u$.

Depending on its properties, a first order phase transition may have interesting implications on the structure of the resulting hybrid stars. The strength of a phase transition is defined by the value of the energy density discontinuity. Notably, if the energy density jump surpasses a critical value ($\Delta\mathcal{E}_{\rm cr}$), then a third family of compact objects may appear on the mass-radius plane. The aforementioned stable branch gives rise to the existence of the so-called twin star solutions~\cite{Gerlach-1968,Kampfer-1981a,Kampfer-1981b,Glendenning-2000,Schertler-2000}. In particular, the strong phase transition (extreme softening) results into an unstable region on the $M-R$ diagram where the mass decreases with increasing central pressure. As a consequence, if the speed of sound in quark matter is sufficient to yield stable configurations for larger pressure, then the resulting stable hybrid branch is going to be disconnected from the hadronic one (shifted to lower radius).
Hence, two stars with an identical mass and a different radius (twin stars) could potentially exist. The value of the critical energy density jump was first studied by Seidovz and it was given by the following expression~\cite{Bielich-2020,Seidov-1971}
\begin{equation} \label{eq14}
    \Delta\mathcal{E}_{\rm cr} = \frac{1}{2}\mathcal{E}_{\rm tr}+\frac{3}{2}P_{\rm tr},
\end{equation}
where $\mathcal{E}_{\rm tr}$ and $P_{\rm tr}$ denote the energy density and pressure of the hadronic phase at the onset of quark deconfinement. As is evident by Eq.~(\ref{eq14}), the larger the value of the transition density, the strongest the phase transition needed for the disconnection of the two stable branches. Notably, the value of $\Delta\mathcal{E}$ acts as a regulator of the radius in the hybrid branch. More precisely, the larger the value of the energy density jump the lower the radius values of the resulting hybrid stars~\cite{Tsaloukidis-2023,Laskos-2023}.

\subsection{Bag constant} \label{2d}
\begin{figure}[h]
  \centering  \includegraphics[width=9 cm,scale=1]{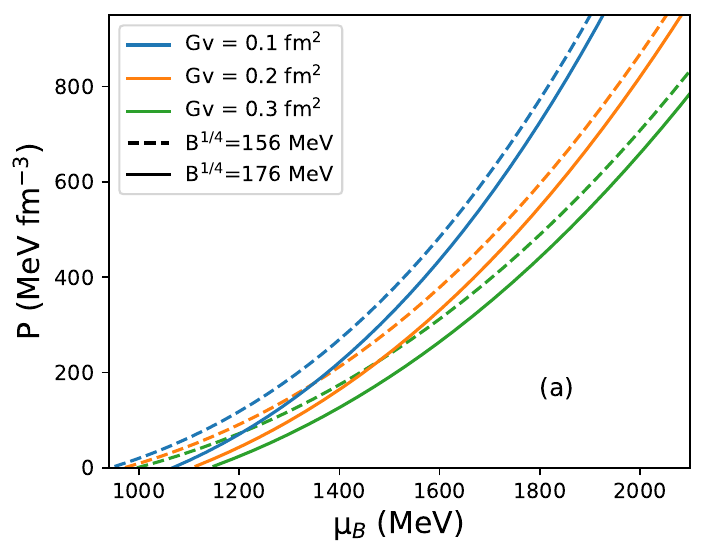} \\
  \includegraphics[width=9 cm,scale=1]{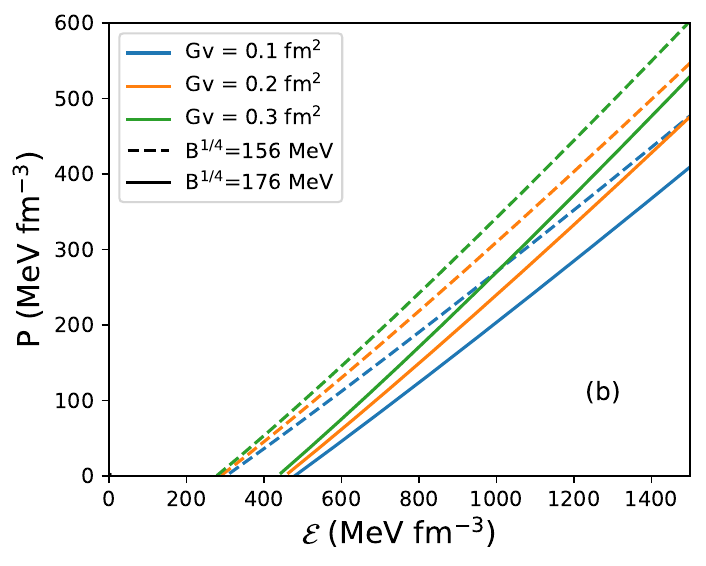} 
  \caption{(a) Pressure as a function of the chemical potential for different values of $G_v$ and $B$. (b) Quark EOSs for different values of $G_v$ and $B$.}
  \label{BcMP}
\end{figure}

We investigate two different scenarios concerning the bag parameter: (a) constant and (b) density dependent. In the first case, considering the values of quark masses fixed ($m_u=m_d=5$ MeV, $m_s=95$ MeV), there are only two free parameters ($G_v$, $B$) that affect the properties of the EOS. In Fig.~\ref{BcMP}(a) we depict the effects of $G_v$ and $B$ on the pressure-chemical potential relation of quark EOSs. We observe that increasing either $G_v$ or $B$ leads to higher values of baryon chemical potential for a given pressure. Thus, quark matter becomes less favorable and the phase transition onset would be shifted to a higher density (for a given hadronic EOS). In Fig.~\ref{BcMP}(b) we illustrate the effects of the aforementioned parameters on the stiffness of the resulting EOSs. As expected, increasing the strength of the repulsive interaction leads to stiffer quark EOSs. In contrast, increasing the bag constant induces softening to the EOS. 

\begin{figure}[h]
  \centering  \includegraphics[width=9 cm,scale=1]{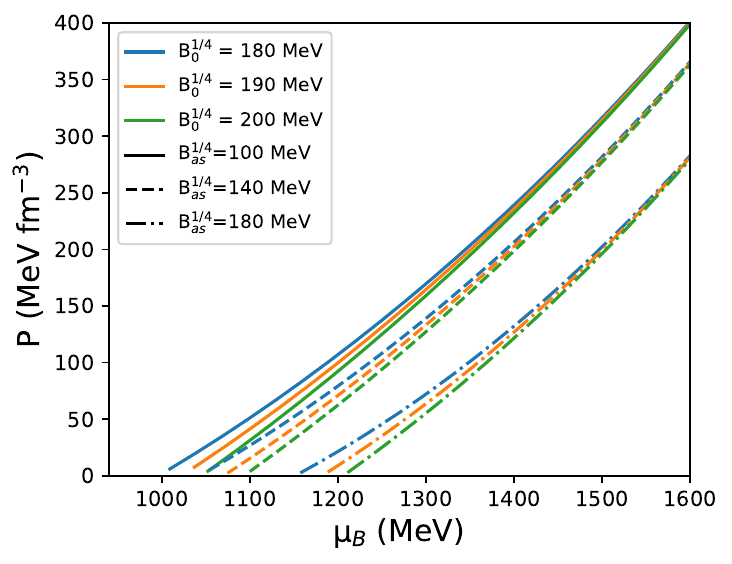} 
  \caption{Pressure as a function of the chemical potential for different values of $B_{as}$ and $B_0$ (the considered values of $G_v$ and 
 $\beta$ are 0.25 fm$^2$ and 0.1, respectively)}. 
  \label{BnMP}
\end{figure}

In order to enrich and extend our study we also examined the scenario of a density dependent bag. Following the discussion of several related studies~\cite{Burgio-2002a,Burgio-2002b,Yazdizadeh-2013,Miyatsu-2015,Sen-2021,Sen-2022,Sen-2023,Pal-2023}, we considered the widely employed Gaussian parametrization 
\begin{equation} \label{eq9}
    B(n) = B_{as}+ (B_0-B_{as}) \exp
    \left[-\beta \left(\frac{n}{n_0}\right)^2\right],
\end{equation}
where $B_0$ and $B_{as}$ stand for the values of $B$  at zero and asymptotically large density. The baryon density in quark matter is given by $n=(n_u+n_d+n_s)/3$, while $n_0$ denotes the nuclear saturation density. It is important to note that in the case of a density dependent $B$ value, we must add an extra term in the quark chemical potentials in order to ensure the thermodynamic consistency. As indicated by Ref.~\cite{Kumar-2023}, in which a vector MIT bag model with a density dependent bag parameter is also employed, the chemical potential will be modified as it follows
\begin{equation} \label{eq12}
    \mu_q=(k_{Fq}^2+m_q^2)^{1/2}+G_v(n_u+n_d+n_s)+\frac{\partial B}{\partial n_q}.
\end{equation}
With regard to the determination of pressure, one may use the standard thermodynamic relations
\begin{equation}
    P_Q=n\frac{d\mathcal{E_Q}}{dn}-\mathcal{E_Q}=\sum_{q} \mu_q n_q - \mathcal{E_Q}.
\end{equation}
In Fig.~\ref{BnMP}, we display the pressure as a function of the baryon chemical potential for different $B_{as}$ and $B_0$ values ($G_v=0.25$ fm$^2$ and $\beta=0.1$). Notably, increasing either one of the parameters leads to a larger chemical potential for a given pressure and hence, the onset of quark deconfinement would be shifted toward  higher densities. It is worth noting that at asymptotically large density all of the curves with equal $B_{as}$ coincide. The latter results from the fact that the density dependent part of the bag term vanishes at high density values. 

\section{Results and discussion} \label{section3}

\subsection{Constant bag parameter}

\begin{figure*}[t]
  \centering  \includegraphics[width=\textwidth,scale=1]{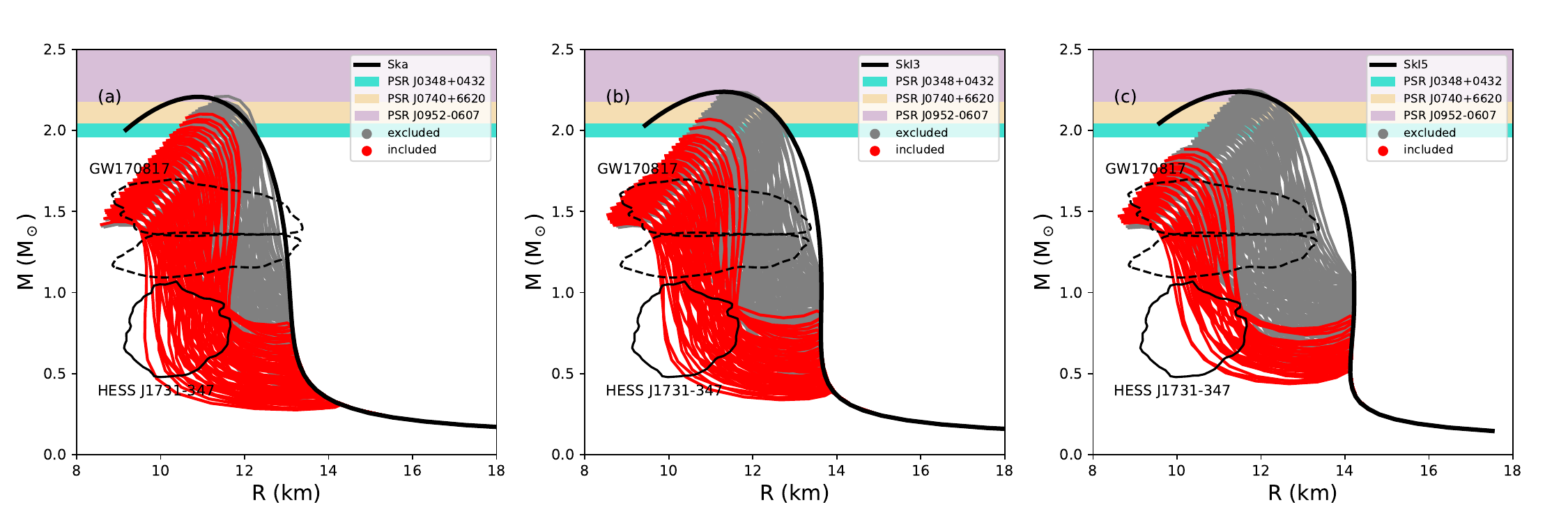}
  \caption{ Mass-radius diagrams for the EOSs constructed with the Skyrme hadronic models and the vector MIT bag quark model with
constant bag pressure. The solid black line stands for the original hadronic EOS. The red (gray) lines stand for the EOSs that are included
(excluded) by the observation of the HESS J1731-347 remnant. The contour regions stand for the $95\%$ confidence interval from
the analysis of the GW170817 event~\cite{Abbott-2017} and the estimation for the mass and radius of the HESS J1731-347 remnant in $1\sigma$~\cite{Doroshenko-2022}. The shaded regions correspond to possible constraints on the maximum mass from the observation of PSR J0348+0432~\cite{Antoniadis-2013}, PSR J0740+6620~\cite{Cromatie-2020} and PSR J0952-0607~\cite{Romani-2022}. Panels (a), (b), and (c) correspond to the cases where the hadronic matter is modeled
via the Ska, SkI3, and SkI5 EOSs, respectively.}
  \label{BcMR}
\end{figure*}

\begin{figure}[h]
  \centering  \includegraphics[width=9 cm,scale=1]{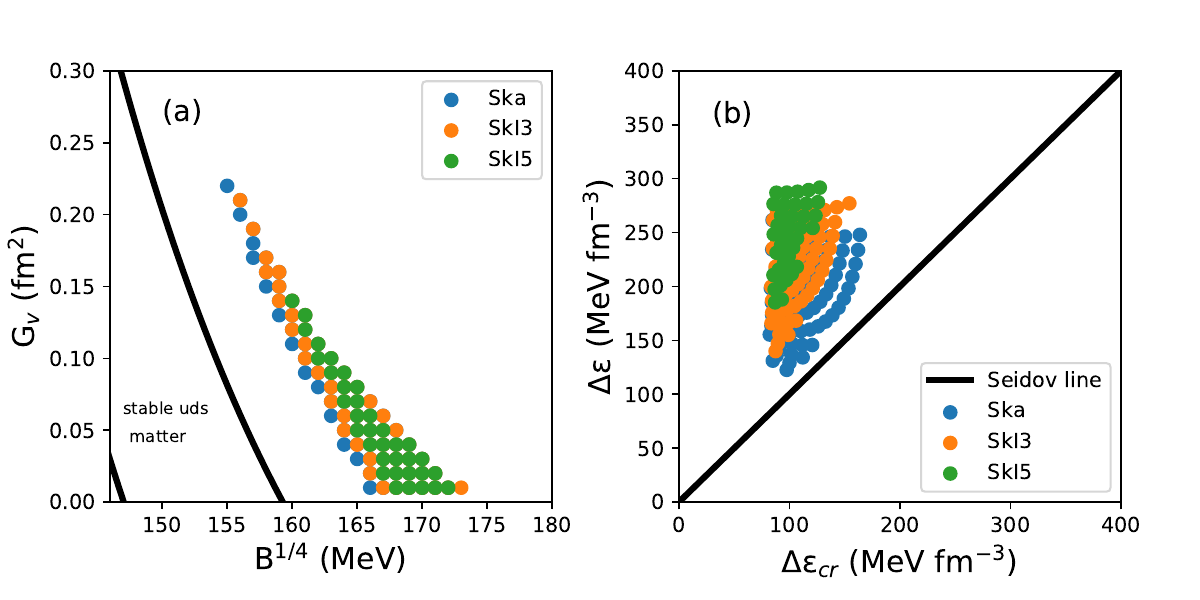}
  \caption{(a) The parametrizations that yield results compatible with HESS J1731-347 ($68\%$ confidence). The axes limits correspond to all of the employed parametrizations. The shaded region correspond to the stability window of uds quark matter. (b) The energy density jump vs the critical energy jump for all of the parametrizations depicted in panel (a).}
  \label{BcParams}
\end{figure}

\begin{figure}[h]
  \centering  \includegraphics[width=9 cm,scale=1]{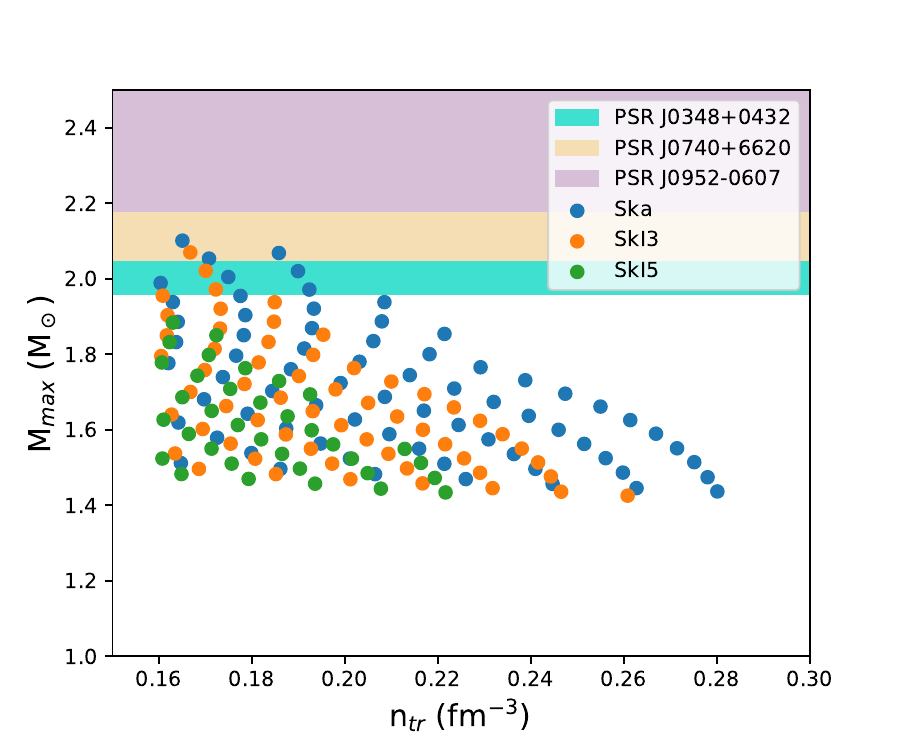}
  \caption{The maximum mass as a function the transition density for each hadronic EOS. The shaded regions correspond to possible constraints on the maximum mass from the observation of PSR J0348+0432~\cite{Antoniadis-2013}, PSR J0740+6620~\cite{Cromatie-2020} and PSR J0952-0607~\cite{Romani-2022}.}
  \label{BcMmaxnt}
\end{figure}

We began our analysis by examining the scenario of a constant bag pressure. In particular, we constructed a large number of quark EOSs, by varying the values of $G_v$ and $B^{1/4}$ in the ranges [0.01-0.3]~fm$^2$ and [146-180]~MeV. Then, we employed the Maxwell construction in order to combine them with the selected low density hadronic EOSs (Ska, SkI3, SkI5). As our main goal was to investigate the compatibility of the resulting hybrid EOSs with the HESS J1731-347 observation, we excluded any model that predicts purely hadronic stars with mass larger than $\sim 1.1M_\odot$ (the latter value exceeds the maximum possible mass of the HESS J1731-347 remnant in $1\sigma$). Furthermore,
the consideration of transition density values below $n_0$ is
not possible, as nuclear matter is known to be stable at that
range. In the absence of a robust theoretical lower limit on the transition density
we set $n_0$ as a loose lower bound, similarly to Ref.~\cite{Zhang-2023}. 

In Fig.~\ref{BcMR} we present the mass-radius dependence for the derived hybrid EOSs. The region defined by the solid contour denotes the mass and radius of the HESS J1731-347 remnant in $1\sigma$~\cite{Doroshenko-2022}. In addition, the areas mapped by the dashed contours correspond to constraints based on the analysis of the GW170817 event ($95\%$ confidence)~\cite{Abbott-2017}. The shaded regions indicate  constraints on the maximum mass from the observation of PSR J0348+0432~\cite{Antoniadis-2013}, PSR J0740+6620~\cite{Cromatie-2020} and PSR J0952-0607~\cite{Romani-2022}. The red curves stand for the EOSs that are compatible with the HESS J1731-347 observation, while the gray ones are excluded. Furthermore, the black solid curves show the $M-R$ relations for the original hadronic EOSs. From Fig.~\ref{BcMR}, considering the hybrid configuration plane in agreement to the HESS J1731-347, we observed that the softening of the hadronic EOSs leads to higher values of maximum neutron star mass. 
Notably, the stiffest hadronic EOS (SkI5) is insufficient to predict hybrid star configurations that surpass the conservative $2M_\odot$ constraint. 

In Fig.~\ref{BcParams}(a) we depict the  quark matter parametrizations that yield results compatible with HESS J1731-347 in $1\sigma$. We observe that as the hadronic EOS stiffens, the area of viable parametrizations decreases. In addition, as $G_v$ increases, the range of possible $B$ values is decreasing. Notably, for a given value of $G_v$, the lower possible value for the bag constant is primarily determined by the fact that all EOSs which predict phase transition below $n_0$ are excluded (increasing either $B$ or $G_v$ shifts the onset of quark deconfinement to higher densities). The highest possible value of $B$ results from an interplay between the predicted transition density and the softening that is induced by the phase transition. In particular, considering the large radii predicted by the stiff hadronic EOSs (see Fig.~\ref{BcMR}) and the low mass and radius of the HESS J1731-347 remnant, one can safely deduce that a strong and early phase transition is essential. The latter is illustrated in Fig.~\ref{BcParams}(b) where we depict the energy density discontinuity as a function of the corresponding critical energy density jump. As is evident, all of the allowed parametrizations yield $\Delta\mathcal{E}\geq\Delta\mathcal{E}_{cr}$, which supports the existence of twin star solutions. Note that, increasing the bag constant would result in an increase of the critical energy density jump (as the phase transition is shifted toward higher densities) and therefore a stronger phase transition would be required for the appearance of a disconnected stable hybrid branch. 

In Table~\ref{tab:table2} we report the maximum possible transition density ($n_{max}$) and the corresponding quark matter parametrization for each of the employed hadronic EOSs. Evidently, as the hadronic EOS becomes stiffer, $n_{max}$ is reduced. Furthermore, the maximum transition density derives from the quark matter parametrization with the lowest employed value of $G_v$ and the highest possible value of $B$. The latter can be understood by considering the following scenario. In principle, increasing the strength of the repulsive interaction stiffens the EOS, while increasing the value of the bag constant leads to softening. Let us now assume that for a given value of $G_v$ the 

\begin{table}[H]
\caption{\label{tab:table2}
The maximum transition density, the corresponding $G_v$ and $B$ values and the predicted maximum mass for each hadronic EOS.}
\begin{ruledtabular}
\begin{tabular}{ccccc}
Model & $n_{max}$ (fm$^{-3}$) & $G_v$ (fm$^{2}$)& $B^{1/4}$ (MeV) &$M_{max}$ ($M_\odot$)\\
\hline
Ska & 0.280 & 0.01 & 172 & 1.44
\\
SkI3 & 0.261 & 0.01 & 173 & 1.43
\\
SkI5 & 0.222 & 0.01 & 172 & 1.43 \\
\end{tabular}
\end{ruledtabular}
\end{table}

\begin{figure*}[t]
  \centering  \includegraphics[width=\textwidth,scale=1]{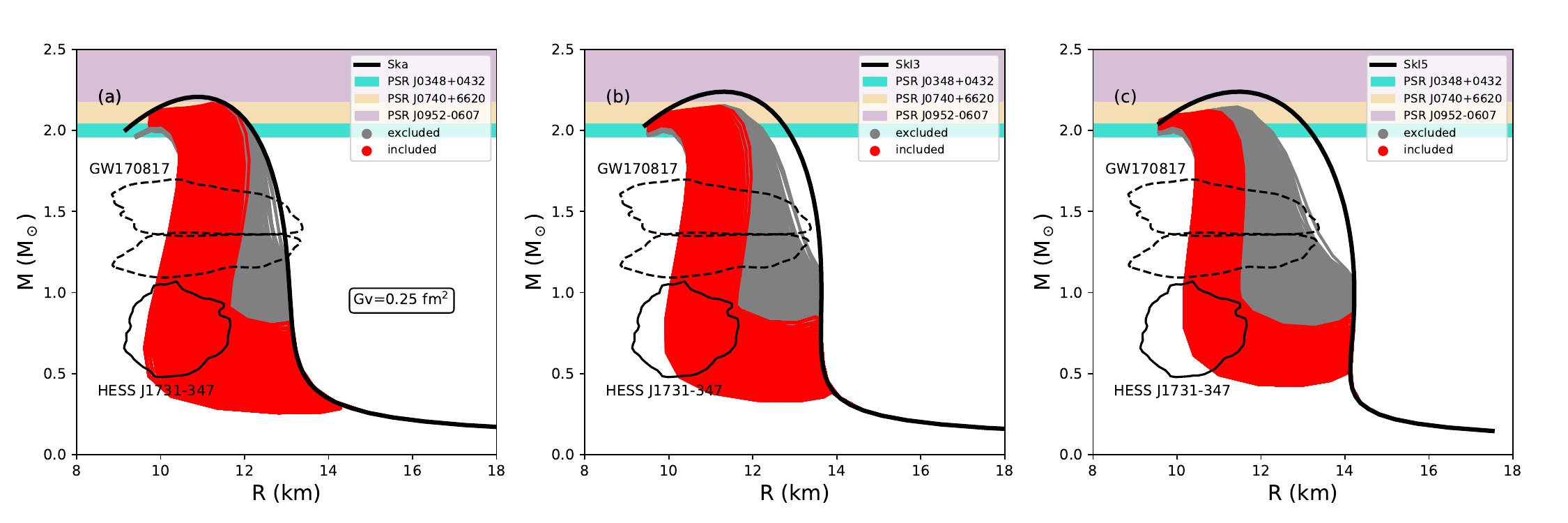}
  \\
    \includegraphics[width=\textwidth,scale=1]{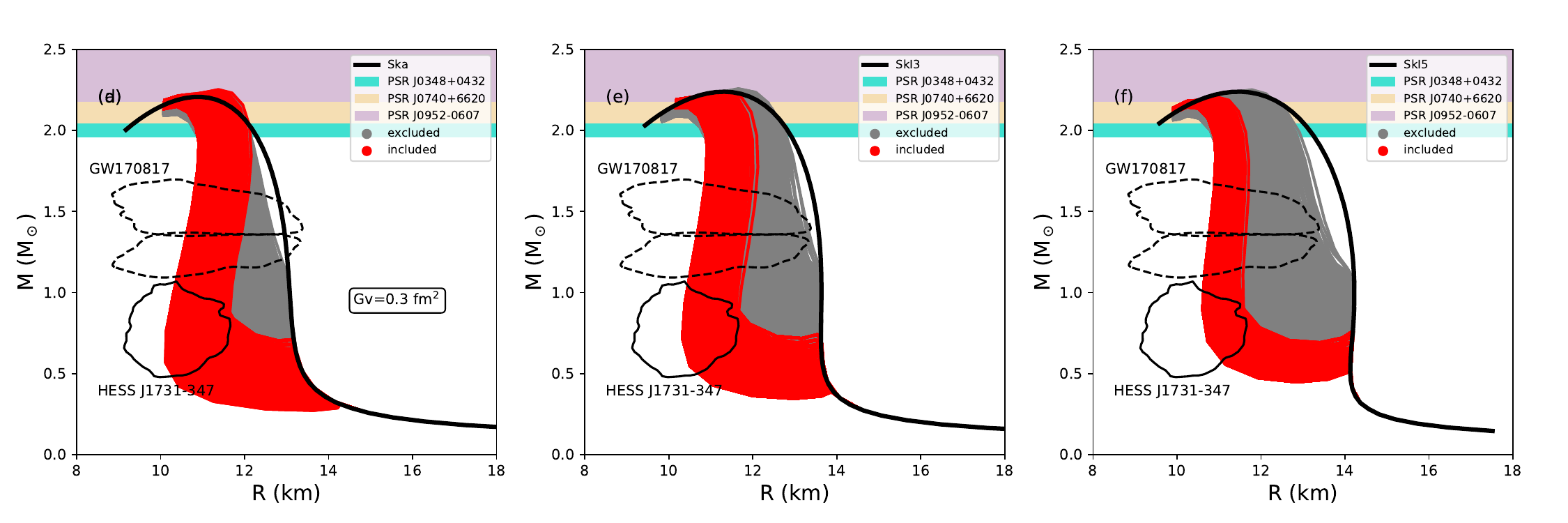}
  \caption{Mass-radius diagrams for the EOSs constructed with the Skyrme hadronic models and the vector MIT bag quark model with
density dependent bag pressure. The solid black line stands for the original hadronic EOS. The red (gray) lines stand for the EOSs that are
included (excluded) by the observation of the HESS J1731-347 remnant. The contour regions stand for the $95\%$ confidence interval from the GW170817 event and the estimation for the mass and radius of the HESS J1731-347 remnant in $1\sigma$. The shaded regions correspond to possible constraints on the maximum mass from the observation of PSR J0348+0432~\cite{Antoniadis-2013}, PSR J0740+6620~\cite{Cromatie-2020} and PSR J0952-0607~\cite{Romani-2022}. The top panels stand for the case of $G_v=0.25$ fm$^2$, while for the bottom panels $G_v=0.3$ fm$^3$. The hadronic matter
is described via the Ska EOS [panels (a),(d)], the SkI3 EOS [panels (b),(e)] and the SkI5 EOS [panels (c),(f)].}
  \label{BnMR}
\end{figure*}

\noindent largest possible value for the explanation of HESS J1371-34 is $B_i$, and the corresponding transition density is $n_i$. If we now increase $G_v$, the only way to obtain a phase transition at $n_i$ is by a employing a value of $B<B_i$. Obviously, the quark EOS in the second case is stiffer as $G_v$ increased and $B$ decreased. As a consequence, the corresponding energy density of quark matter at the hadron-quark interface is going to be smaller (and $\Delta\mathcal{E}$ will be reduced), since a stiffer EOS requires lower energy density for supporting the same amount of pressure. Considering that $\Delta\mathcal{E}$ regulates the radius dislocation between the hadronic and the hybrid branch (lower $\Delta\mathcal{E}$ results in higher radii in the hybrid branch)~\cite{Tsaloukidis-2023,Laskos-2023}, then the stiffer EOS is not going to satisfy the mass and radius constraints for the HESS J1731-347 remnant. Thus, increasing $G_v$ leads to viable phase transitions only for $n<n_i$ and hence, the maximum transition density derives from the softest allowed parametrization (maximum $B$ and minimum $G_v$). The latter analysis also explains our previous observation that beyond a value of $G_v$ it is impossible to obtain hybrid configurations compatible with HESS J1731-347. In particular, as $G_v$ increases, the largest possible value for the transition density decreases. It is worth noting that at some point the maximum transition density reaches $n_0$ and therefore further increasing of $G_v$  would result in phase transitions below nuclear saturation density.

While softening the quark EOS allows for an increase of the maximum possible transition density, the maximum mass of the predicted compact stars is decreased . As it is evident from Table~\ref{tab:table2} the reported parametrizations significantly fail to produce stable massive stars. In Fig.~\ref{BcMmaxnt} we present the maximum mass of all EOSs that agree with the HESS J1731-347 observation as a function of the corresponding transition density. We observed that the conservative $2M_\odot$ limit places stringent constraints on the onset of quark deconfinement. For the Ska EOS, the observation of PSR J0348+0432 restricts the phase transition onset below $\sim$ 0.2 fm$^{-3}$. However, as the hadronic EOS stiffens, the latter value gets lower. Considering that a phase transition close to nuclear saturation is unrealistic, we may conclude that the model of a constant bag pressure gets strongly disfavored as the symmetry energy slope increases.

\subsection{Density dependent bag parameter}

We continued our study by analyzing the case of a density dependent bag pressure. In this scenario we considered two different values for the strength of the interaction. In particular, we set $G_v$ to $0.25$~fm$^2$ and $0.3$~fm$^2$. Then, we fixed $\beta$ to a typical value of 0.1 and we varied the values of $B_{as}^{1/4}$ and $B_{0}^{1/4}$ in the ranges $[0, 200]$ MeV and $[B_{as}^{1/4}, B_{as}^{1/4}+200]$ MeV, respectively. Notably, for the two limiting cases $\{B_{as}^{1/4}=0,B_0^{1/4}=200 $ ${\rm MeV}\}$ and $\{ B_{as}^{1/4}=200,B_0^{1/4}=200 $ ${\rm MeV}\}$  the predicted transition densities allow for purely hadronic stars with mass higher than $\sim1.1M_\odot$. Therefore, we may safely deduce that it is impossible to obtain viable hybrid configurations beyond these parametrizations (as increasing $B_{as}^{1/4}$ or $B_{0}^{1/4}$ shifts the phase transition toward higher densities). 

In Fig.~\ref{BnMR} we depict the mass-radius relations for  the constructed hybrid EOSs. The top panels correspond to the case where $G_v=0.25$ fm$^2$, while the bottom panels to $G_v=0.3$ fm$^2$ (the drawn astrophysical constraints are the same as in Fig.~\ref{BcMR}).  Once again, the red curves denote the cases where the HESS J1731-347 constraint is satisfied, while the gray ones are excluded. Interestingly, all of the configurations that agree with the observation of the HESS J1731-347 can also support massive compact stars that exceed the $2M_\odot$ limit. In addition, we observed that as the strength of the repulsive interaction increases the maximum possible mass for the hybrid configurations becomes larger. However, the disconnection of the hybrid branch appeared to reach lower radii for the softest scenario ($G_v=0.25$ fm$^2$).

As shown in Tables~\ref{tab:table3} and \ref{tab:table4}, the maximum mass for the EOSs with the maximum possible transition density is not strongly affected by the employed hadronic model. However, as the hadronic EOS stiffens, the corresponding transition density decreases. Furthermore, stiffening the quark EOS leads to a reduction of $n_{max}$ for a given hadronic model. Therefore, the highest value for the onset of quark deconfinement derives from the combination of the softest possible models of hadronic and quark matter. We conclude that information on the maximum mass of compact stars in combination with a robust lower limit on the transition density may provide important insight concerning the stiffness of the two phases.

\begin{figure}
  \centering  \includegraphics[width=9 cm,scale=1]{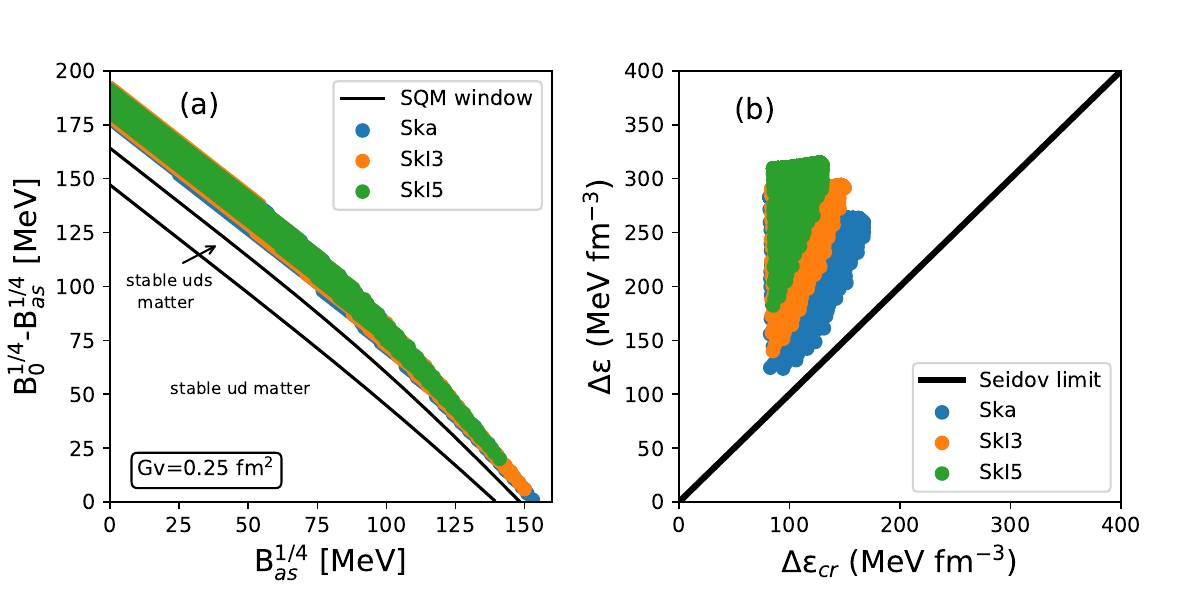} 

  \caption{(a) The parametrizations that yield results compatible with HESS J1731-347 ($G_v=0.25$ fm$^2$). The shaded region corresponds to the stability window of uds quark matter. (b) The energy density jump vs the critical energy jump for all of the parametrizations depicted in panel (a).}
  \label{BnParams1}
\end{figure}

\begin{figure}
  \centering  \includegraphics[width=9 cm,scale=1]{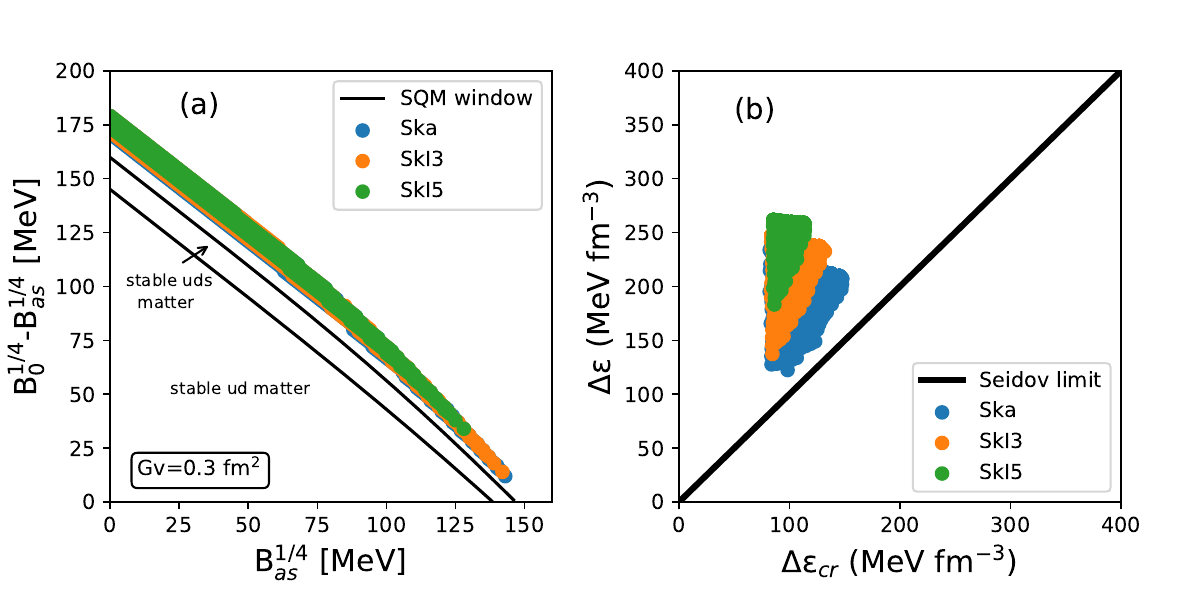} 

  \caption{same as Fig.~\ref{BnParams1} but for $G_v=0.3$ fm$^2$}
  \label{BnParams2}
\end{figure}

In Fig.~\ref{BnParams1}(a) and Fig.~\ref{BnParams2}(a) we depict all of the allowed quark matter parametrizations for the cases of $G_v=0.25$ and $0.3$ fm$^2$, respectively. Similarly to the case of the constant bag parameter, we observed that the stiffness of the hadronic phase does not affect the location of the parameter space. However, it affects its width. In particular, the stiffer the hadronic EOS, the fewer the allowed parametrizations. Figs.~~\ref{BnParams1}(b) and ~\ref{BnParams2}(b) illustrate the expected necessity for a strong phase transition. It is worth commenting that the lower value for the coupling constant allows for significantly stronger phase transitions (larger $\Delta\mathcal{E}$).

\begin{table}
\caption{\label{tab:table3}
The maximum transition density in the case of $G_v=0.25$ fm$^2$ and the corresponding transition pressure and maximum mass configuration.}
\begin{ruledtabular}
\begin{tabular}{cccc}
Model & $n_{max}$ (fm$^{-3}$)& $p_{max}$ (MeV fm$^{-3}$)  &$M_{max}$ ($M_\odot$)\\
\hline
Ska & 0.285 & 18.667  & 2.035 \\
SkI3 & 0.256 & 16.761  & 2.031\\
SkI5 & 0.225 & 13.597  & 2.036\\
\end{tabular}
\end{ruledtabular}
\end{table}

\begin{table}
\caption{\label{tab:table4}
The maximum transition density in the case of $G_v=0.3$ fm$^2$ and the corresponding transition pressure and maximum mass configuration }
\begin{ruledtabular}
\begin{tabular}{cccc}
Model & $n_{max}$ (fm$^{-3}$)& $p_{max}$ (MeV fm$^{-3}$)  &$M_{max}$ ($M_\odot$)\\
\hline
Ska  & 0.259 & 14.323  & 2.147 \\
SkI3 & 0.232 & 12.854 &  2.140 \\
SkI5 & 0.203 & 10.442  & 2.145 \\
\end{tabular}
\end{ruledtabular}
\end{table}

\begin{figure}[h]
  \centering  \includegraphics[width=9 cm,scale=1]{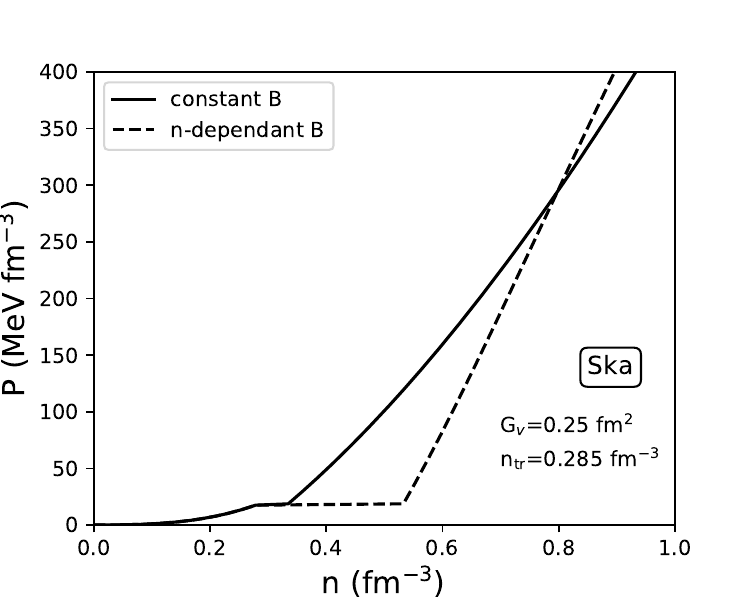}
  \caption{Hybrid EOSs with a phase transition at $n_{\rm tr}=0.285$ fm$^{-3}$. The solid (dashed) curve denotes the case of a constant (density dependent) bag. }
  \label{strongPT}
\end{figure}

Contrary to the previous case (constant $B$), we observed that the resulting hybrid EOSs satisfy the conservative $2M_\odot$ constraint for larger values of the transition density. As we have shown in Fig.~\ref{BcParams}(a), it is not possible to obtain a hybrid EOS that is compatible with HESS J1731-347 for a quark EOS with constant $B$ and $G_v=0.25$ fm$^2$. The latter results from the fact that the quark EOS is very stiff and hence the resulting energy density jump is small. In Fig.~\ref{strongPT}, we depict two hybrid EOSs (constant and density dependent bag), which are characterized by same value of $G_v=0.25$ fm$^2$ and predict the same transition density (the low density phase is described by the Ska model). For the scenario of the $n-$dependent bag pressure we have used the EOS with the maximum transition density (see Table~\ref{tab:table3}).  As it is illustrated, the introduction of an $n-$dependent bag allows for the desired soft behavior at low baryon density, leading to a significant density discontinuity. The latter is a direct consequence of the density dependence, as an extra negative term appears in the baryon chemical potential. In particular, if $B$ is constant then the baryon chemical potential is given by
\begin{equation} \label{eq15}
    \mu_B^c=\mu_u+2\mu_d=\sum_{q}\sqrt{k_{Fq}^2+m_q^2}+9G_v n,
\end{equation}
where the index $q$ denotes the u,d,s quark flavors. However, if $B$ is density dependent then
\begin{equation} \label{eq16}
%\begin{split}  
\mu_B^n=\sum_{q}\sqrt{k_{Fq}^2+m_q^2}+9G_v n+\sum_{q}\frac{\partial B}{\partial n_q},
%\end{split}
\end{equation}
The first two terms in Eqs.~(\ref{eq15}) and (\ref{eq16}) are identical for a given baryon density. The latter is fairly simple to understand since the extra term in Eq.~(\ref{eq12}) is the same regardless of the quark flavor. Therefore, the expressions for chemical equilibrium and charge neutrality (see Eqs.~(\ref{eq7}) and (\ref{eq8})), which determine the quark fractions for a given baryon density, remain identical (either for constant or $n-$dependent $B$). As a consequence, the chemical potential difference for the two cases results from the bag related term. It is also straightforward that the latter term is negative.~Let us now consider the two different scenarios depicted in Fig.~\ref{strongPT}. At the point of the phase transition it is obvious that $\mu_B^c(n_{Q_1})=\mu_B^n(n_{Q_2})$, where $n_{Q_1}$ and $n_{Q_2}$ stand for the baryon densities of the two quark models. Since the two first terms of Eqs.~(\ref{eq15}) and (\ref{eq16}) are identical for a given baryon density, then it is obvious that $n_{Q_2}> n_{Q_1}$. Therefore, the density jump in the $n-$dependent case is always going to be wider (for a given transition density).

\begin{figure}
  \centering  \includegraphics[width=9 cm,scale=1]{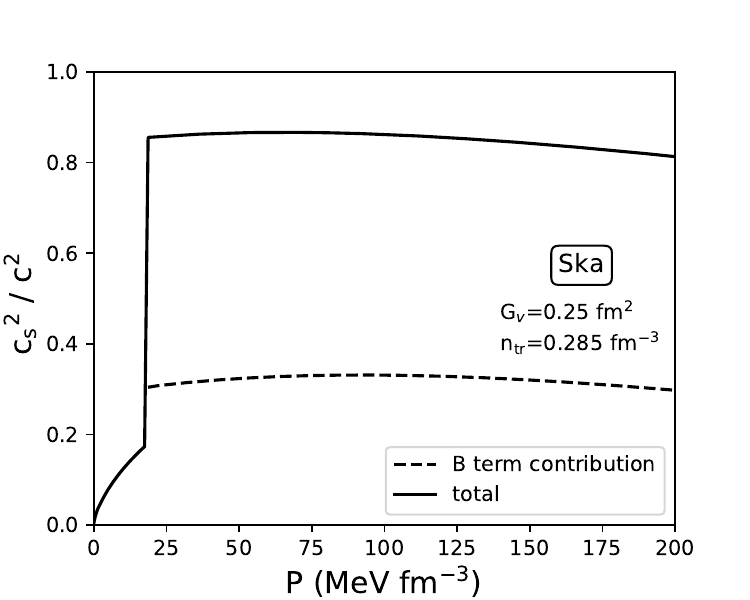}
  \caption{ The contribution of the bag related term (in the
pressure function) to the squared speed of sound (in units of
squared speed of light $c^2$) for the hybrid EOS with the largest
possible transition density for $G_v=0.25$ fm$^2$.}
  \label{soundspeed}
\end{figure}

Despite the softening induced by the introduction of a density dependent bag pressure, the resulting hybrid EOSs can support massive compact stars. The reason is twofold. Firstly, as we discussed, the introduction of an $n-$dependent bag  allows for a soft EOS behavior at low density even for larger $G_v$ values (stronger repulsive interaction/larger speed of sound $c_s/c=\sqrt{dP/d\mathcal{E}}$).  Secondly, the induced softening vanishes at higher baryon density. The latter effect manifests through the contribution of the bag related term to the total speed of sound. In Fig.~\ref{soundspeed}, we depict the aforementioned contribution for the EOS with the maximum transition density in the case of $G_v=0.25$~fm$^2$. Interestingly, more than $35\%$ of the squared speed of sound
results from the density dependent bag term in the pressure
function. We need to highlight that the difference between $B_0$ and $B_{as}$ plays a crucial role concerning the size of the contribution. If it is large enough it may lead to violation of causality. Therefore, one needs to pay special attention when a density dependent bag model is employed.

\subsection{PSR J0030+0451}

The NICER constraints on the mass and radius of PSR J0030+0451 have provided an important tool in the quest of unraveling the nature of dense nuclear matter~\cite{Miller-2019,Riley-2019,Raaijmakers-2019}. At this point we wish to examine the compatibility of the resulting hybrid EOSs, discussed in the previous section,~with the NICER information on PSR J0030+0451.~According to the analysis of Miller {\it et al.}~\cite{Miller-2019}, the mass and radius of PSR J0030+0451   are equal to $M=1.44^{+0.15}
_{-0.14} M_\odot$ and $R= 13.02^{+1.24}_
{-1.06}$km (in $68\%$ confidence or $1\sigma$). Notably, a relevant analysis has been conducted in Ref.~\cite{Riley-2019}, providing rather similar results. Considering the low radius predicted by the constructed hybrid EOSs, we aim to examine the implications of PSR J0030+0451 by employing the lowest estimation on its radius (in different confidence levels). In the present study, we employ the results of Ref.~\cite{Miller-2019}, as the predicted lowest possible radius is slightly larger (compared to the analysis of Ref.~\cite{Riley-2019}), providing a more challenging constraint on the derived hybrid EOSs. A similar approach has been previously employed in Ref.~\cite{Christian-2020}, where the authors utilized the mass and radius of PSR J0030+0451 (within $1\sigma$) in order to impose constraints on the onset of strong phase transitions.
\begin{figure}[t]
  
  \includegraphics[width=9 cm,scale=1]{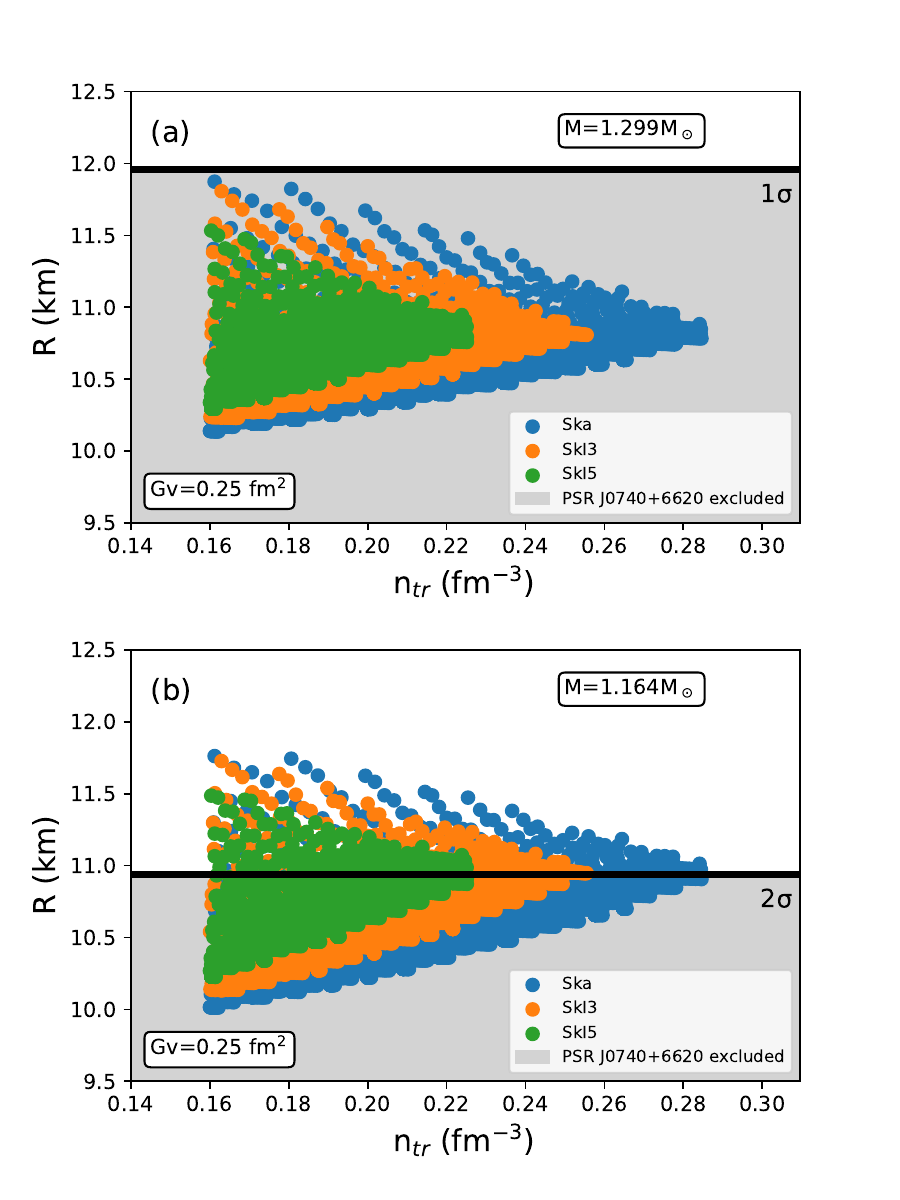}
  \caption{(a) The radius predicted from the resulting parametrizations and the $1\sigma$ constraint from NICER. (b) The radius predicted from the resulting parametrizations and the $2\sigma$ constraint from NICER. The value of the coupling constant is equal to $G_v=0.25$ fm$^2$.}
  \label{Nicer1}
\end{figure}
\begin{figure}[t]
  
  \includegraphics[width=9 cm,scale=1]{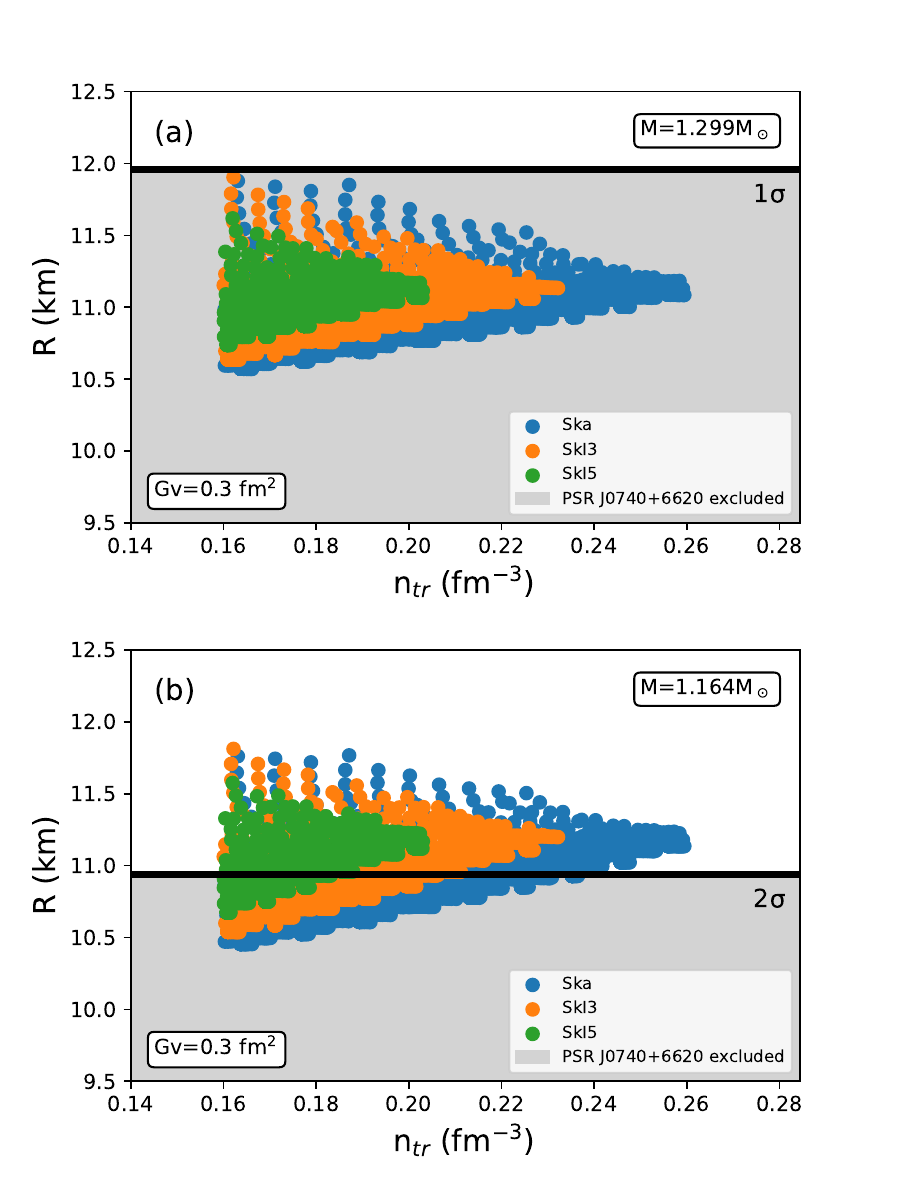}
  \caption{Same as Fig.~\ref{Nicer1} but for the case of $G_v=0.3$ fm$^2$.}
  \label{Nicer2}
\end{figure}

In Fig.~\ref{Nicer1} we depict the radius of $1.299M_\odot$ and $1.164M_\odot$ hybrid stars (in the case of $G_v=0.25$ fm$^2$) as a function of the corresponding transition density. The aforementioned values correspond to the lowest possible mass of PSR J0030+0451 in $1\sigma$ and $2\sigma$, respectively~\cite{Miller-2019}. Fig.~\ref{Nicer2} is the same as Fig.~\ref{Nicer1}, but for the case where the coupling constant is taken to be $G_v=0.3$ fm$^2$. Our results indicate that, it is very difficult to simultaneously explain PREX-II, HESS J1731-347 and PSR J0030+0451 in the framework of hybrid stars. In particular, none of the constructed hybrid EOSs satisfies the PSR J0030+0451 constraints in $1\sigma$. Note that this result is in agreement with the relevant analysis conducted in Ref.~\cite{Christian-2020}, where the authors suggested that the mass and radius of PSR J0030+0451 disfavor strong phase transitions below $\sim1.7n_0$. We conclude that, if the estimations of PREX-II, HESS J1731-347 and PSR J0030+0451 are verified, within $1\sigma$, the scenario of purely hadronic or hybrid stars would be strongly challenged. As consequence the scenario of strange quark stars or the possible admixture of dark matter in compact stars would be favored. As one can observe in the bottom panels of Figs.~\ref{Nicer1} and \ref{Nicer2}, there are hybrid configurations that are marginally compatible with the PSR J0030+0451 in $2\sigma$, for all of the possible transition density values. However, as the transition density increases the compatibility becomes more marginal. Interestingly, the analysis of Ref.~\cite{Christian-2020} suggests that the $2\sigma$ estimations on PSR J0030+0451 should exclude any phase transition below $\sim1.4n_0$, which is in contrast with our results. However, note that the authors of the aforementioned study considered strong phase transition only for the case $\Delta\mathcal{E}=350$ MeV. According to our analysis the explanation of HESS J1731-347 requires $\Delta\mathcal{E}\lesssim320$ MeV (see Figs. \ref{BnParams1}(b) and \ref{BnParams2}(b)). Larger $\Delta\mathcal{E}$ values could be produced by considering lower $G_v$ values. However, this would potentially result in contradiction with the observation of massive compact stars. Finally, one may argue that since the radius of hybrid configurations may increase as a function of the mass, then the study of configurations based on the lowest mass estimation for the PSR J0030+0451 may exclude EOSs that are in fact viable. However, based on the derived 2D contour on the mass-radius plane for PSR J0030+0451, higher mass configurations would have more stringent radius constraints [see Fig.~7(b) in Ref.~\cite{Miller-2019}]. Therefore, the study of higher mass compact stars would not alter the conclusion that the agreement between our hybrid EOSs and PSR J0030+0451 (in $2\sigma$) is only marginal.

\subsection{PSR J0952-0607} \label{3d}

Pulsar PSR J0952-0607, discovered by Bassa
{\it et al.}~\cite{Bassa-2017}, has a frequency of $f = 709$ Hz, making it the
fastest known spinning pulsar in the disk of the Milky Way.
Recently, Romani {\it et al.}~\cite{Romani-2022} reported that PSR
J0952-0607 has a mass of $M=2.35\pm 0.17M_\odot$, which is
the largest well measured mass found to date. The existence of such a massive compact object has triggered the exploration of possible constraints on the
nuclear EOS and it is likely to revise many of
the theoretical predictions concerning the basic properties
of neutron stars~\cite{Tsaloukidis-2023,Ecker-2023}. In particular, the extreme softening induced by a strong phase transition has important implications on the resulting maximum mass. According to Ref.~\cite{Christian-2021}, the observation of supermassive pulsars may
exclude the possibility of twin star solutions. However, our results clearly demonstrate that a strong phase transition is essential for the explanation of the HESS J1731-347 remnant (assuming a stiff low density phase). Considering the tension between strong phase transitions and massive compact stars, we aim to examine the compatibility between the derived hybrid EOSs and the reported mass of PSR J0952-0607.

\begin{figure}[t]
  \centering  \includegraphics[width=9 cm,scale=1]{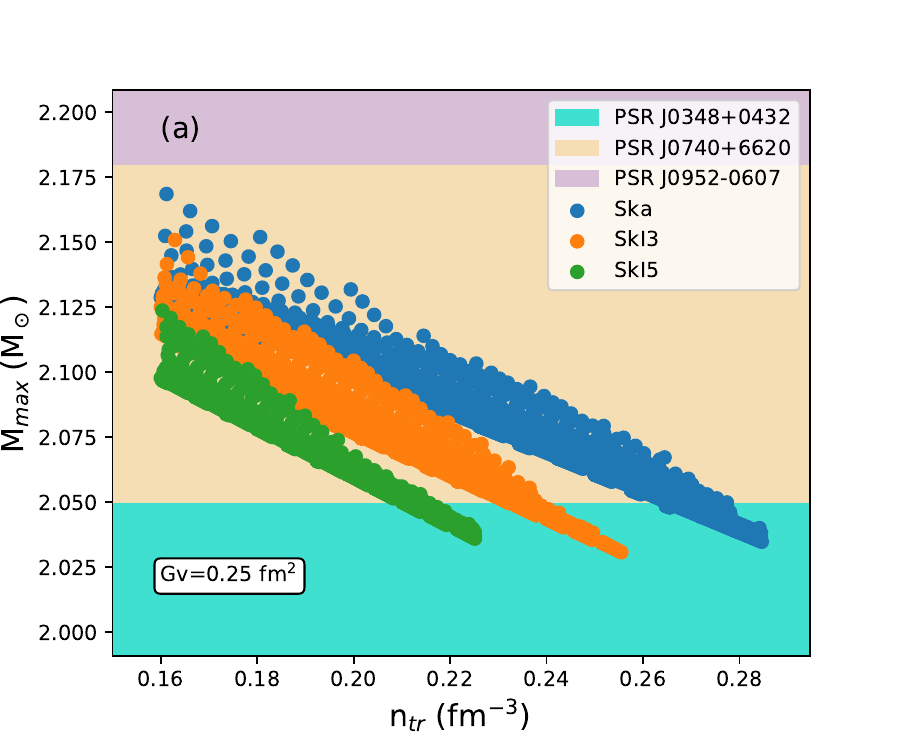}
  \\
  \includegraphics[width=9 cm,scale=1]{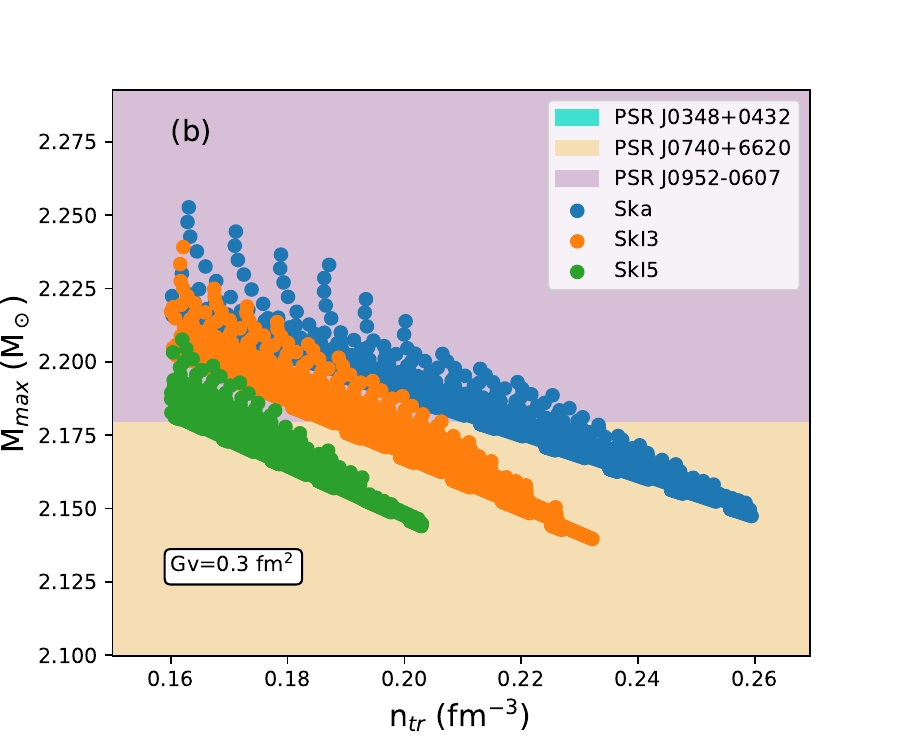}
  \caption{The maximum mass as a function the transition density for each hadronic EOS. The shaded regions correspond to possible constraints on the maximum mass from the observation of PSR J0348+0432~\cite{Antoniadis-2013}, PSR J0740+6620~\cite{Cromatie-2020} and PSR J0952-0607~\cite{Romani-2022}. Panels (a) and (b) depict the cases where $G_v=0.25$ and $G_v=0.3$ fm$^2$.}
  \label{gaussmmaxn}
\end{figure}

\begin{table}
  \caption{The maximum mass of hybrid stars rotating at 709 Hz for different values of the transition density and different hadronic EOSs. The value of the coupling constant is equal to $G_v=0.25$ fm$^2$. The values of $n_{max}$ for this case can be found in Table~\ref{tab:table3}. The missing values correspond to the fact that no phase transition occurs.}
  \label{tab:table5}
  \begin{tblr}{Xccc} 
    \hline
    \hline
     $n_{tr}$ (fm$^{-3}$) & \SetCell[c=3]{} $M_{max}$ ($M_\odot$) &&\\
    \hline
     & Ska & SkI3 & SkI5 \\
    \hline
    0.20 & 2.190 & 2.151 & 2.120 \\
    0.24 & 2.147 & 2.105 & - \\
    0.28 & 2.095 & - & - \\
    $n_{max}$ & 2.076 & 2.073 & 2.080 \\
   \hline
   \hline
  \end{tblr}
\end{table}

\begin{table}
  \caption{The maximum mass of hybrid stars rotating at 709 Hz for different values of the transition density and different hadronic EOSs. The value of the coupling constant is equal to $G_v=0.3$ fm$^2$. The values of $n_{max}$ for this case can be found in Table~\ref{tab:table4}. The missing values correspond to the fact that no phase transition occurs.}
  \label{tab:table6}
  \begin{tblr}{Xccc} 
    \hline
    \hline
     $n_{tr}$ (fm$^{-3}$) & \SetCell[c=3]{} $M_{max}$ ($M_\odot$) &&\\
    \hline
     & Ska & SkI3 & SkI5 \\
    \hline
   0.20 & 2.276 & 2.243 & 2.201 \\
   0.22 & 2.248 & 2.211 & - \\
   0.24 & 2.225 & -     & - \\
   $n_{max}$ & 2.194 & 2.190 & 2.196 \\
   \hline
   \hline
  \end{tblr}
\end{table}

In Fig.~\ref{gaussmmaxn} we depict the maximum mass for all of the hybrid EOSs that are compatible with the observation of HESS J1731-347 as a function of the transition density. As one can observe, the earlier the phase transition, the larger the corresponding maximum mass. Notably, the observation of PSR J0952-0607 places very stringent constraints on the coupling constant. In particular, for $G_v=0.25$ fm$^2$ none of the EOSs can support stars with  mass above 2.18$M_\odot$. In contrast, for $G_v=0.3$ fm$^2$ we find that several configurations satisfy the aforementioned mass constraint. Interestingly, the information on PSR J0952-0607 allows for the imposition of constraints on the transition density. In particular, considering the softest hadronic model (Ska), the phase transition has to occur for a baryon density below $\sim0.23$~fm$^{-3}$. The latter value reduces as the stiffness of the hadronic phase increases.

\begin{figure*}[t]
  \centering  \includegraphics[width=\textwidth,scale=1]{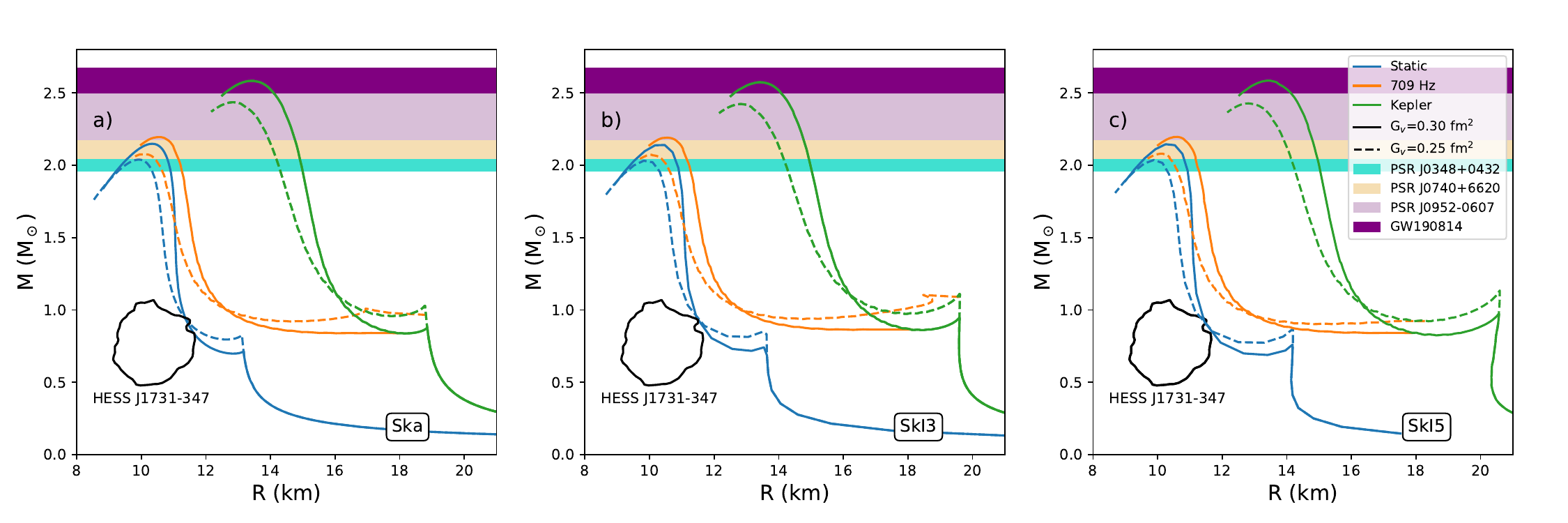}

  \caption{Mass-radius dependence for static and rapidly rotating compact stars constructed with the EOSs that predict the maximum transition densities (see Tables~\ref{tab:table3} and~\ref{tab:table4}) . The solid (dashed) curves correspond to the case of $G_v=0.3$ fm$^2$ ($G_v=0.25$ fm$^2$). The solid contour region stands for the mass and radius of the HESS J1731-347 remnant in $1\sigma$. The shaded regions correspond to possible constraints on the maximum mass from the observation of PSR J0348+0432~\cite{Antoniadis-2013}, PSR J0740+6620~\cite{Cromatie-2020}, PSR J0952-0607~\cite{Romani-2022} and GW190814~\cite{Abbott-2020}. Panels (a), (b) and (c) correspond to the cases where the hadronic matter is modeled via the Ska, SkI3, and SkI5
EOSs, respectively.}
  \label{GW190814}
\end{figure*}

The aforementioned results address the compatibility between the derived hybrid EOSs and the existence of a pulsar with $M=2.35\pm 0.17 M_\odot$, under the assumption that the massive compact object is static (and hence spherically symmetric). However, PSR J0952-0607 is one of the fastest rotating pulsars ever detected. It is noteworthy that fast rotation may have an effect on the resulting maximum mass, as the centrifugal force contributes to the battle against gravity. For the latter reason we revisit our previous calculation by introducing the effects of rotation. Specifically, we selected the EOSs which for a given value of transition density (rounded to two decimals) predict the maximum possible mass. Then, we constructed rotating equilibrium stellar models at rotational frequency $f=709$~Hz. In Tables~\ref{tab:table5} and \ref{tab:table6} we report the maximum mass of compact stars, rotating at 709 Hz, for different values of the transition density. As shown in Table~\ref{tab:table5} the inclusion of rotation enables the explanation of PSR J0952-0607 mass even for $G_v=0.25$~fm$^2$. However, this holds only for the softest hadronic model and the transition density is constrained below $\sim 0.2$ fm$^{-3}$. For the case of $G_v=0.3$ fm$^2$, the constructed hybrid EOSs are compatible with the mass of PSR J0952-0607 for all of the possible transition density values.  Interestingly, the EOS with the maximum transition ($\sim 1.7 n_0$) density barely satisfies the PSR J0952-0607 constraints. Note that decreasing $G_v$ would increase the maximum transition density but it would also decrease the corresponding maximum mass. In addition, increasing $G_v$ would increase the maximum mass but it would reduce the maximum transition density. Based on the softest model at hand (Ska), our results indicate that the existence of HESS J1731-347 and PSR J0952-0607 suggests a strong phase transition below $\sim 1.7 n_0$.

\subsection{GW190814}

In August 2019, the LIGO/Virgo Collaboration reported the observation of a compact binary coalescence involving a $22.2-24.3M_\odot$ black hole and a compact object with a mass of $2.5-2.67M_\odot$~\cite{Abbott-2020}. Notably, the low mass component  falls into the neutron star–black hole mass gap and therefore it resembles either the most massive neutron star or the lightest black hole ever detected. Subsequently, the possibility of such a massive neutron star has triggered a scientific debate concerning the nature of the secondary component in the GW190814 event. 

It is worth mentioning that none of the hybrid EOSs derived in the present study agrees with the lowest predicted limit for the mass of the unidentified compact object in GW190814. However, as we discussed in the previous section, the consideration of rapid rotation will increase the maximum possible mass of a neutron star, compared to a nonrotating one. Considering the extreme scenario in the softening of the EOS, that is essential for the explanation of the HESS J1731-347 remnant, we aim to examine the possibility of explaining the existence of a $2.5-2.67M_\odot$ hybrid star under the assumption of extreme rotation. In Fig.~\ref{GW190814} we depict the mass-radius dependence of the hybrid EOSs that predict the latest phase transitions in the cases of $G_v=0.25$ and $0.3$ fm$^2$. Apart from the static case, we also include the Keplerian sequence (i.e., stars rotating at the mass shedding limit). For completeness we also depict the sequence of stars rotating at 709 Hz (discussed in the previous Sec.~\ref{3d}). In the case of $G_v=0.3$ fm$^2$, the consideration of fast rotation yields results that are compatible with the existence of a $2.5-2.67 M_\odot$ compact star. Taking into account that this holds for the EOS with the lowest maximum mass for $G_v=0.3$ fm$^2$, we may conclude that all the configurations that agree with HESS J1731-347 are also in accordance with the constraints from GW190814, considering the rapid rotation of neutron stars (since higher maximum mass configurations in static neutron stars lead to higher maximum mass configurations in maximally rotating ones~\cite{Largani-2022}).
However, in the case of $G_v=0.25$ fm$^2$ the sequence of maximally rotating compact stars cannot reach $2.5M_\odot$. Therefore, it would be necessary to consider configurations with earlier phase transitions in order to explain the GW190814 event.

One final remark needs to be made concerning the effect of rotation on the radius difference of twin stars. By examining the case of rotation at 709 Hz one may safely deduce that the inclusion of rotational effects may result in a significantly wider unstable region on the mass-radius plane and hence a larger radius difference between twin stars. More precisely, in the case of $G_v=0.25$~fm$^2$ the possible radius difference between the twin configurations is sufficiently larger compared to the static case. This may be of particular importance concerning the possibility of exploiting inertial oscillation modes in order to probe the hadron-quark phase transition~\cite{Laskos-2023}. For instance, it has been shown that the radius of a compact star has a snotable effect on the so-called $r$-mode instability window, which essentially determines the limiting spin frequency for stable $r$-mode oscillations~\cite{Papazoglou-2016}. Finally, it is important to comment that as the considered rotational frequency increases, the sequences of stellar models will eventually terminate on the hybrid branch of the Keplerian sequence. The latter implies the possible elimination of stable twin star configurations beyond a value of rotational frequency.

For the numerical integration of the equilibrium equations we used the publicly available RNS code by Stergioulas
and Friedman~\cite{Stergioulas-1995}; this code is based on the method developed by Komatsu, Eriguchi and Hachisu~\cite{Komatsu-1989} and modifications introduced by Cook, Shapiro and Teukolsky~\cite{Cook-1994}).

\section{Conclusion}\label{section4}

In the present study we focused on the explanation of the compact object in the HESS J1731-347 remnant assuming a stiff low density hadronic EOS, which is favored by the recent PREX-II experiment. In particular, we considered three widely employed Skyrme effective interactions and we have combined them with a vector MIT bag model in order obtain a variety of hybrid EOSs. Finally, we investigated two distinct scenarios concerning the energy density of the bag. In the first case it was taken to be a constant, whereas in the second case we adopted a widely used Gaussian parametrization. It was found that the explanation of HESS J1731-347 requires a strong phase transition (which leads to the appearance of a disconnected hybrid branch) in both cases.

With regard to the scenario of the constant bag parameter, we found that the conservative $\sim2M_\odot$ limit for the maximum mass places stringent constraints on the derived hybrid EOSs. More precisely, the onset of quark deconfinement for the softest employed hadronic model is restricted below $\sim 0.2$ fm$^{-3}$. In addition, stiffening the hadronic phase shifts the onset of the phase transition to even lower densities. Considering that a phase transition close to nuclear saturation density is unrealistic we conclude that the model of a constant bag pressure gets strongly disfavored as the low density phase stiffens. 

In the case of the density dependent bag parameter we were able to show that it is possible to explain the observation of the HESS J1731-347 remnant simultaneously with the existence of massive compact stars, even for larger values of the transition density. The latter derives from the fact that the introduction of density dependence modifies the baryon chemical potential in a way that early and strong phase transitions are allowed for larger values of the repulsive interaction coupling constant. As a consequence, the quark EOS may exhibit a soft behavior at low baryon density (allowing for phase transition with large $\Delta\mathcal{E}$) and then stiffen in order to support stable massive compact stars. The direct effect of a density dependent bag parameter in the stiffening of the EOS manifests through its contribution to the total speed of sound.

Apart from the HESS J1371-374 constraints we also considered the recent analysis for the mass and radius of PSR~J0030+0451. We found that it is very difficult  to simultaneously explain the aforementioned observations within $1\sigma$ in the context of purely hadronic or hybrid stars. As a consequence, the scenarios of purely quark stars or compact stars that are admixed with dark matter may be favored. Interestingly, we were able to show that the derived hybrid EOSs are compatible with PSR~J0030+0451 in $2\sigma$ for all of the possible transition density values. However, we need to comment that as the transition density increases the agreement becomes more marginal.

Taking into account the essential extreme EOS softening that is required for the explanation of the HESS J1731-347 remnant, we also considered the possible limits on the maximum mass of compact stars based on the observation PSR~J0952-060 ($2.35 \pm 0.17 M_\odot$). We demonstrated that the explanation of such a massive compact object requires a sufficiently large value for the coupling constant of the repulsive interaction. For the largest employed $G_v$ value we showed that the EOS with the maximum transition density barely reaches $2.18M_\odot$. Considering that increasing the coupling constant would result in earlier phase transitions and also that decreasing it would lower the corresponding maximum mass we were able to constrain the onset of quark deconfinement below $\sim1.7 n_0$.

Finally, we paid particular attention to the possible interpretation of the supermassive compact object involved in the GW190814 event. Considering static stellar configurations, we illustrated that it is not possible to obtain stable $\sim2.5 M_\odot$ compact stars based on the derived hybrid EOSs. However, the introduction of rapid rotation enabled the explanation of such a massive compact object. In particular, our results indicate that HESS J1731-347 and GW190814 are potentially compatible under the assumption that the unidentified component in the latter gravitational event is a maximally rotating hybrid star.
\\
\\
\\
\noindent {\it Note added.$-$}Shortly after the submission of our paper to the journal, a preprint has appeared that also studies the properties of hybrid stars in light of recent astronomical data by considering various configurations for the stiffness of the EOS~\cite{Li-2023n}. Contrary to our study, which attempts to provide some insight on the phenomenological parameters of the quark EOS (interaction, bag parameter), the authors conduct an alternative analysis based on a constant speed of sound model.

\section*{ACKNOWLEDGMENTS}
P. S. K. acknowledges the support of the Special Account
for Research Funds (ELKE) of the Aristotle University of
Thessaloniki in the framework of the Research Fellowships
for Postdoctoral Researchers (Fellowship Project: 50186,
Fellowship Number: 673402).


\begin{thebibliography}{99}
%%%%%%%%%%%%%%%%%%%%%%%%%%%%%%%%%%%%%%%%%%%%%%%%%%%%%
\bibitem{Baldo-2016}M. Baldo, G. F. Burgio, Part. Nucl. Phys. {\bf 91}, 203-258 (2016).
\bibitem{Roca-Maza-2018} X. Roca-Maza and N. Paar, Prog.
Part. Nucl. Phys. {\bf 101}, 96 (2018).
\bibitem{Reed-2021} B.T. Reed, F.J. Fattoyev , C.J. Horowitz ,and J. Piekarewicz, Phys. Rev. Lett. {\bf 126}, 172503 (2021) 
\bibitem{Abrahamyan-2012} S. Abrahamyan, Z. Ahmed, H. Albataineh, K. Aniol, D. S.
Armstrong et al., Phys. Rev. Lett. {\bf 108}, 112502 (2012).
\bibitem{Horowitz-2012} C.J. Horowitz, Z. Ahmed, C. M. Jen, A. Rakhman, P. A.
Souder et al., Phys. Rev. C {\bf 85}, 032501(R) (2012).
\bibitem{Adhikari-2021} D. Adhikari et al. (PREX collaboration), Phys. Rev. Lett. {\bf 126},
172502 (2021)
\bibitem{Hebeler-2013}  K. Hebeler, J. Lattimer, C. Pethick, and A. Schwenk,
Astrophys. J. {\bf 773}, 11 (2013).
\bibitem{Zhang-2013} Z. Zhang and L.-W. Chen, Phys. Lett. B {\bf 726}, 234 (2013).
\bibitem{Hagen-2016} G. Hagen et al., Nat. Phys. {\bf 12}, 186 (2016).
\bibitem{Drischler-2020} C. Drischler, R. J. Furnstahl, J. A. Melendez, and D. R.
Phillips, Phys. Rev. Lett. {\bf 125}, 202702 (2020).
\bibitem{Horowitz-2014} C. J. Horowitz, E. F. Brown, Y. Kim, W. G. Lynch, R.
Michaels, A. Ono, J. Piekarewicz, M. B. Tsang, and H. H.
Wolter, J. Phys. G {\bf 41}, 093001 (2014).
\bibitem{Adhikari-2022}D. Adhikari et al. (CREX Collaboration), Phys. Rev. Lett. {\bf 129}, 042501 (2022).
\bibitem{Reinhard-2022} P.G. Reinhard, X. Roca-Maza, W. Nazarewicz, Phys. Rev. Lett. {\bf 129}, 232501 (2023).
\bibitem{Tagami-2022} Shingo Tagami, Tomotsugu Wakasa, Masanobu Yahiro, Results Phys. {\bf 43}, 106037 (2022).
\bibitem{Miyatsu-2023} Tsuyoshi Miyatsu, Myung-Ki Cheoun, Kyungsik Kim , Koichi Saito, Phys. Lett. {\bf 843B}, 138013 (2023).
\bibitem{mKumar-2023} Mukul Kumar, Sunil Kumar, Virender Thakur, Raj Kumar, B. K. Agrawal, and Shashi K. Dhiman,
Phys. Rev. C {\bf 107}, 055801 (2023).
\bibitem{Reed-2023} Reed, B.T.; Fattoyev, F.J.; Horowitz, C.J.; Piekarewicz, arXiv:2305.19376 [nucl-th] (2023).
\bibitem{Becker-2018} D. Becker {\it et al.}, Eur. Phys. J. A {\bf 54}, 208 (2018) 
\bibitem{Abbott-2018} B.P. Abbott et al. (LIGO Scientific Collaboration and Virgo Collaboration)
Phys. Rev. Lett. {\bf 121}, 161101 (2018).
\bibitem{Abbott-2017} B.P. Abbott et al. (LIGO Scientific Collaboration and Virgo Collaboration)
Phys. Rev. Lett. {\bf 119}, 161101 (2017).
\bibitem{Thakur-2023} Virender Thakur, Raj Kumar, Pankaj Kumar, Mukul Kumar, C. Mondal, Kaixuan Huang, Jinniu Hu, B. K. Agrawal, and Shashi K. Dhiman, Phys. Rev. C {\bf 107}, 015803 (2023).
\bibitem{Essick-2021a} Reed Essick, Ingo Tews, Philippe Landry, and Achim Schwenk,
Phys. Rev. Lett. {\bf 127}, 192701 (2021).
\bibitem{Essick-2021b} Reed Essick, Philippe Landry, Achim Schwenk, and Ingo Tews, Phys. Rev. C {\bf 104}, 065804 (2021).
\bibitem{Yeunhwan-2022} L. Yeunhwan, and W. Jeremy Holt, Galaxies {\bf 10(5)}, 99 (2022)
\bibitem{Thapa-2023} Vivek Baruah Thapa, and Monika Sinha, EPJ Web of Conferences {\bf 279}, 10003 (2023).
\bibitem{Soares-2023} Bruno A. de Moura Soares, César H. Lenzi, Odilon Lourenço, Mariana Dutra,
Month. Not. Roy. Astron. Soc. {\bf 525(3)}, 4347–4357 (2023).
\bibitem{Chen-2023} Manjia Chen, Dawei Guan, Chongji Jiang, Junchen Pei, arXiv:2309.11245 [nucl-th] (2023).
\bibitem{Thapa-2022} Vivek Baruah Thapa, and Monika Sinha	arXiv:2203.02272 [nucl-th] (2022).
\bibitem{Sarkar-2023} Trisha Sarkar, Vivek Baruah Thapa, and Monika Sinha
Phys. Rev. C {\bf 108}, 035801 (2023).
\bibitem{Doroshenko-2022} V. Doroshenko, V. Suleimanov, G. Phlhofer, and Andrea
Santangelo, Nat. Astron. {\bf 6}, 1444 (2022).
\bibitem{Alford-2023} V. J. A. J. Alford and J. P. Halpern,  Astrophys. J. {\bf944}, 36 (2023).
\bibitem{DiClemente-2023} Francesco Di Clemente, Alessandro Drago, Giuseppe Pagliara, arXiv:2211.07485 [astro-ph.HE] (2023).
\bibitem{Suwa-2018} Y. Suwa, T. Yoshida, M. Shibata, H. Umeda, and K. Takahashi, Mon. Not. Roy. Astron. Soc. {\bf481}, 3305 (2018).
\bibitem{Horvath-2023} J. E. Horvath, L. S. Roch, L. M. de Sá1, P. H. R. S. Moraes, L. G. Barão1, M. G. B. de Avellar, A. Bernardo and R. R. A. Bachega, Astron. Astrophys. {\bf672}, L11 (2023).
\bibitem{Oikonomou-2023} P.T. Oikonomou and Ch.C. Moustakidis
Phys. Rev. D {\bf 108}, 063010 (2023).
\bibitem{Das-2023} H. C. Das, Luiz L. Lopes, Mon. Not. Roy. Astron. Soc. {\bf525}, 3571 (2023).
\bibitem{Rather-2023} Ishfaq A. Rather, Grigoris Panotopoulos, and Ilidio Lopes, arXiv:2307.03703 [astro-ph.HE] (2023).
\bibitem{Tsaloukidis-2023} Lazaros Tsaloukidis, P.S. Koliogiannis, A. Kanakis-Pegios, and Ch.C. Moustakidis
Phys. Rev. D {\bf 107}, 023012 (2023).
\bibitem{Brodie-2023} L. Brodie and A. Haber
Phys. Rev. C {\bf 108}, 025806 (2023).
\bibitem{Sagun-2023} Violetta Sagun, Edoardo Giangrandi, Tim Dietrich, Oleksii Ivanytskyi, Rodrigo Negreiros,
and Constanca Providencia, Astrophys. J. {\bf 958}, 49 2023 (2023).
\bibitem{Huang-2023} Kaixuan Huang, Hong Shen, Jinniu Hu, Ying Zhang, arXiv:2306.04992v1 [nucl-th] (2023).
\bibitem{Li-2023} Jia Jie Li, Armen Sedrakian, Phys Lett. {\bf 844B}, 138062 (2023).
\bibitem{Kubis-2023} Sebastian Kubis, Wlodzimierz Wojcik, David Alvarez Castillo, and Noemi Zabari, arXiv:2307.02979 [nucl-th] (2023).
\bibitem{Routaray-2023} Pinku Routaray, H. C. Das, Jeet Amrit Pattnaik, Bharat Kumar, arXiv:2307.12748 [math.NA] (2023)
\bibitem{Kohler-1976} H.S. Köhler, Nucl. Phys. A {\bf 258}, 301 (1976).
\bibitem{Reinhard-1995} P.-G. Reinhard, and H. Flocard, Nucl. Phys. A {\bf584}, 467-488 (1995).
\bibitem{Danielewicz-2009} P. Danielewicz, and J. Lee, Nucl. Phys. A {\bf818}, 36 (2009)
\bibitem{Baym-1971} G. Baym; P. Pethick, and P. Sutherland, Astrophys. J, {\bf 170},  299 (1971).
\bibitem{Chabanat-1997} E. Chabanat, P. Bonche, P. Haensel, J. Meyer, and R. Schaeffer, Nucl. Phys. A {\bf 627}, 710 (1997).
\bibitem{Costantinou-2014} Constantinos Constantinou, Brian Muccioli, Madappa Prakash, and James M. Lattimer
Phys. Rev. C {\bf89}, 065802 (2014).
\bibitem{Costantinou-2015} Constantinos Constantinou, Brian Muccioli, Madappa Prakash, James M. Lattimer, Ann. Phys. {\bf 363}, 533-555 (2015).
\bibitem{Tsaloukidis-2019} L. Tsaloukidis, Ch. Margaritis, and Ch. C. Moustakidis
Phys. Rev. C {\bf99}, 015803 (2019).
\bibitem{Margaritis-2021} Ch. Margaritis, P. S. Koliogiannis, A. Kanakis-Pegios, and Ch. C. Moustakidis
Phys. Rev. C {\bf 104}, 025805 (2021).
\bibitem{Biswas-2022} Bhaskar Biswas, Astrophys. J. {\bf 926}, 75 (2022)
\bibitem{Landry-2022} Philippe Landry, Kabir Chakravarti, arXiv:2212.09733 [astro-ph.HE] (2022).
\bibitem{Mikheev-2023} S. Mikheev, D. Lanskoy, A.
Nasakin, T. Tretyakova, Particles {\bf6}, 847–863 (2023).
\bibitem{Shlomo-2006} S. Shlomo, V. M. Kolomietz, and G. Colò. The EPJ A {\bf30(1)}, 23–30 (2006).
\bibitem{Colo-2008} G. Colò. Phys. Part. Nucl. {\bf39(2)}, 286–305 (2008).
\bibitem{Garg-2018} U. Garg and G. Colò, Prog. Part. Nucl. Phys {\bf101}, 55-95 (2018).
\bibitem{Stone-2014} J. R. Stone, N. J. Stone, and S. A. Moszkowski
Phys. Rev. C {\bf89}, 044316 (2014).
\bibitem{compose} https://compose.obspm.fr/
\bibitem{Gulminelli-2015} F. Gulminelli, and Ad. R. Raduta, Phys. Rev. C {\bf 92}, 055803 (2015)
\bibitem{Baym-1976} G. Baym, and S.A. Chin, Phys. Lett. {\bf 62(2)B}, 241-244 (1976).
\bibitem{Lopes-2021a} Luiz L. Lopes, Carline Biesdorf, and Débora P Menezes, Phys. Scr. {\bf 96}, 065303 (2021).
\bibitem{Lopes-2021b} Luiz L. Lopes, Carline Biesdorf, K D Marquez, and Débora P. Menezes, Phys. Scr. {\bf 96}, 065302 (2021).
\bibitem{Klahn-2015} T. Klähn and T. Fischer, Astrophys. J. {\bf810}, 134 (2015).
\bibitem{Gomez-2016} B. Franzon, R. Gomes, and S. Schramm, Mon. Not. Roy.
Astron. Soc. {\bf463}, 571 (2016).
\bibitem{Gomez-2019a} R. Gomes, P. Char, and S. Schramm, Astrophys. J. {\bf877}, 139
(2019).
\bibitem{Gomes-2019b} R. Gomes, V. Dexheimer, S. Han, and S. Schramm, Mon. Not. Roy. Astron. Soc. {\bf485}, 4873
(2019).
\bibitem{Jaikumar-2021} P. Jaikumar, A. Semposki, M. Prakash, and C. Constantinou
Phys. Rev. D {\bf 103}, 123009 (2021).
\bibitem{Costantinou-2021} C. Constantinou, S. Han, P. Jaikumar, and M. Prakash
Phys. Rev. D {\bf104}, 123032 (2021)
\bibitem{Zhao-2022} T. Zhao, C. Constantinou, P. Jaikumar, and M.Prakash
Phys. Rev. D {\bf 105}, 103025 (2022).
\bibitem{Costantinou-2023} C, Constantinou, T. Zhao, S. Han, and M. Prakash
Phys. Rev. D {\bf 107}, 074013 (2023).
\bibitem{Lyra-2023} F. Lyra, L. Moreira, R. Negreiros, R. O. Gomes, and V. Dexheimer, Phys. Rev. C {\bf 107}, 025806 (2023).
\bibitem{Kumar-2022} A. Kumar, V. B. Thapa, and M. Sinha, Mon. Not. R. Astron.
Soc. {\bf 513}, 3788 (2022).
\bibitem{Kumar-2023} A. Kumar, V. B. Thapa, and M. Sinha
Phys. Rev. D {\bf 107}, 063024 (2023).
\bibitem{Serot-1992} B.D. Serot, Rep. Prog. Phys. {\bf 55}, 1855 (1992).
\bibitem{Yang-2021} Shu-Hua Yang, Chun-Mei Pi, Xiao-Ping Zheng, and Fridolin Weber
Phys. Rev. D {\bf 103}, 043012  (2021).
\bibitem{Yang-2023} Shu-Hua Yang, Chun-Mei Pi, Xiao-Ping Zheng, and Fridolin Weber, Universe {\bf9}, 202 (2023).
\bibitem{Yasutake-2016} N. Yasutake, H. Chen, T Maruyama, and T. Tatsumi, N J. Phys.: Conf. Ser. {\bf 665} 012068 (2016).
\bibitem{Mariani-2017} M. Mariani, M. Orsaria,  and H. Vucetich, Astron. Astrophys. {\bf 601 }  A21 (2017).
\bibitem{Bielich-2020} J. Schaffner-Bielich, {\it Compact Star Physics} (Cambridge University Press, Cambridge, England, 2020).
\bibitem{Gerlach-1968}U. H. Gerlach, Phys. Rev. {\bf172}, 1325 (1968).
\bibitem{Kampfer-1981a} J. Kampfer, Phys. A: Math. Gen. {\bf14}, L471 (1981).
\bibitem{Kampfer-1981b} J. Kampfer, Phys. Lett. {\bf101B}, 366 (1981).
\bibitem{Glendenning-2000} N. K. Glendenning and C. Kettner, Astron. Astrophys.
{\bf353}, L9 (2000).
\bibitem{Schertler-2000} K. Schertler, C. Greiner, and J. Schaffner-Bielich, and M. H. Thoma, Nucl. Phys. {\bf A677}, 463 (2000).
\bibitem{Seidov-1971}  Z. F. Seidov, Sov. Astron. {\bf15}, 347 (1971).
\bibitem{Laskos-2023} P. Laskos-Patkos and Ch. C. Moustakidis,
Phys. Rev. D {\bf107}, 123023 (2023).
\bibitem{Burgio-2002a} G. F. Burgio, M. Baldo, P. K. Sahu, and H.-J. Schulze
Phys. Rev. C {\bf 66}, 025802 (2002).
\bibitem{Burgio-2002b} G. F. Burgio, M. Baldo, P. K. Sahu, A.B. Santra, and H.-J. Schulze, Phys. Lett.  {\bf 526B}, 19–26 (2002).
\bibitem{Miyatsu-2015} Tsuyoshi Miyatsu, Myung-Ki Cheoun, and Koichi Saito, Astrophys. J. {\bf 813}, 135 (2015).
\bibitem{Yazdizadeh-2013} T. Yazdizadeh, and G. H. Bordbar 
Astrophysics {\bf 56}, 121–129 (2013).
\bibitem{Sen-2021} Debashree Sen, Naosad Alam, and Gargi Chaudhuri, J. Phys. G: Nucl. Part. Phys. {\bf48}, 105201 (2021).
\bibitem{Sen-2022} Debashree Sen, Naosad Alam, and Gargi Chaudhuri
Phys. Rev. D {\bf 106}, 083008 (2022).
\bibitem{Sen-2023} Suman Pal, Soumen Podder, Debashree Sen, and Gargi Chaudhuri
Phys. Rev. D {\bf 107}, 063019 (2023).
\bibitem{Pal-2023} Suman Pal and Gargi Chaudhuri
Phys. Rev. D {\bf 108}, 103028 (2023).
\bibitem{Zhang-2023} Nai-Bo Zhang and Bao-An Li,
Phys. Rev. C {\bf108}, 025803 (2023).
\bibitem{Antoniadis-2013} J. Antoniadis, P. Freire, N. Wex et al., Science {\bf 340}, 448
(2013).
\bibitem{Cromatie-2020} H. Cromartie, E. Fonseca, S. Ransom et al., Nat. Astron. {\bf 4},
72 (2020). 
\bibitem{Romani-2022} R. G. Romani, D. Kandel, A. V. Filippenko, T. G. Brink, and W. Zheng, Astrophys. J. Lett. {\bf 934}, L17 (2022).
\bibitem{Miller-2019} M. C. Miller {\it et al.}, Astrophys. J. Lett {\bf887}, L24 (2019).
\bibitem{Riley-2019} T. E. Riley {\it et al.}, Astrophys. J. Lett {\bf887}, L21 (2019).
\bibitem{Raaijmakers-2019} G. Raaijmakers {\it et al.}, Astrophys. J. Lett {\bf887}, L24 (2019).
\bibitem{Christian-2020} Jan-Erik Christian and Jürgen Schaffner-Bielich, Astrophys. J. Lett. {\bf 894}, L8 (2020).
\bibitem{Bassa-2017}  C. G. Bassa, Z. Pleunis, J. W. T. Hessels et al., Astrophys. J.
Lett. {\bf846}, L20 (2017).
\bibitem{Ecker-2023} C. Ecker and L. Rezzolla, Mon. Not. Roy. Astron. Soc. {\bf519}, 2615–2622 (2023).
\bibitem{Christian-2021} J. E. Christian and J. Schaffner-Bielich, Phys. Rev. D {\bf 103},
063042 (2021).
\bibitem{Abbott-2020} R. Abbott, Astrophys. J. Lett. {\bf 896}, L44 (2020).
\bibitem{Largani-2022} Noshad Khosravi Largani, Tobias Fischer, Armen Sedrakian, Mateusz Cierniak, David E. Alvarez-Castillo, and David B. Blaschke, Mon. Not. Roy. Astron. Soc. {\bf 515}, 3539–3554 (2022).
\bibitem{Papazoglou-2016} M. C. Papazoglou and C. C. Moustakidis, Astrophys. Space Sci. {\bf 361}, 98 (2016).
\bibitem{Stergioulas-1995} N. Stergioulas and J.L. Friedman, Astrophys. J. \textbf{444}, 306 (1995).
\bibitem{Komatsu-1989} H. Komatsu, Y. Eriguchi and I. Hachisu, Mon. Not. Roy. Astron. Soc. \textbf{237}, 355 (1989).
\bibitem{Cook-1994} G.B. Cook, S.L. Shapiro and S.A. Teukolsky, Astrophys. J. \textbf{422}, 227 (1994).
\bibitem{Li-2023n} Jia Jie Li, Armen Sedrakian, and Mark Alford 	arXiv:2401.02198 [astro-ph.HE] (2023).
\end{thebibliography}
\end{document}